%% file: main.tex
\documentclass[english, 11pt, letterpaper]{article}
\usepackage{amsmath,amssymb,amsthm}
\usepackage[authoryear,round]{natbib}
\usepackage[nodisplayskipstretch]{setspace}
\usepackage{bm}
\usepackage{threeparttable}
\usepackage{graphicx}
\usepackage{enumerate}
\usepackage{commath}			
\usepackage{bibentry}
\usepackage{booktabs}
\usepackage[titletoc,title]{appendix}
\usepackage{multirow}
\usepackage{tabularx}
\usepackage{dsfont}
\usepackage{rotating} 
\usepackage[font={small}]{subcaption}
\usepackage[left=1in, right=1in, top=1in, bottom=1in]{geometry}
\usepackage{float}
\PassOptionsToPackage{table}{xcolor}
\usepackage[colorinlistoftodos,textsize=small]{todonotes}
\usepackage{xcolor}
\usepackage[ruled,vlined, linesnumbered]{algorithm2e}
\usepackage{booktabs}
\usepackage{threeparttable}
\usepackage{tikz}
\usetikzlibrary{shapes.geometric}

\allowdisplaybreaks

\usepackage{url}
\usepackage{xr-hyper}  
\usepackage[
	breaklinks,
	colorlinks=true,
	pagebackref=false,
	pdfauthor={},%
	pdftitle={},%
	citecolor=blue,
	linkcolor=blue,
	urlcolor=blue
	]{hyperref}      

\newtheoremstyle{remark2}{1ex}{1ex}%
      {}
      {}
      {\bf}
      {.}
      {5pt}
      {\thmname{#1}\thmnumber{ #2}\thmnote{ \slshape{(#3)}}} 
\theoremstyle{remark2}

\newtheoremstyle{remark3}{1ex}{1ex}%
      {\it}
      {}
      {\bfseries}
      {.}
      {5pt}
      {\thmname{#1}\thmnumber{ #2*}\textnormal{\thmnote{ (#3)}}} 
\theoremstyle{remark3}

\usepackage{babel}
\input{ee.sty}

\@ifundefined{bibfont}{  } 
    {  }



\newcommand{\Comments}{1}
\newcommand{\mynote}[2]{\ifnum\Comments=1\textcolor{#1}{#2}\fi}
\newcommand{\mytodo}[2]{\ifnum\Comments=1%
  \todo[linecolor=#1!80!black,backgroundcolor=#1,bordercolor=#1!80!black]{#2}\fi}

\begin{document}

\baselineskip18pt
\renewcommand\floatpagefraction{.9}
\renewcommand\topfraction{.9}
\renewcommand\bottomfraction{.9}
\renewcommand\textfraction{.1}
\setcounter{totalnumber}{50}
\setcounter{topnumber}{50}
\setcounter{bottomnumber}{50}
\abovedisplayskip1.5ex plus1ex minus1ex
\belowdisplayskip1.5ex plus1ex minus1ex
\abovedisplayshortskip1.5ex plus1ex minus1ex
\belowdisplayshortskip1.5ex plus1ex minus1ex

\title{Online Conditional Vine Copulas: Forecasting Electricity Demand\thanks{
    The authors would like to thank Álvaro Rodriguez for excellent research assistance. 
    Simon Hirsch is thankful for the support and funding received by Statkraft Trading GmbH. This work represents the authors opinions and not necessarily reflects Statkraft's position.
    Christoph Hanck, Christian Jobelius Schulz, and Florian Ziel gratefully acknowledge financial support of the Deutsche Forschungsgemeinschaft (DFG, German Research Foundation) within TRR 391: Spatio-temporal Statistics for the Transition of Energy and Transport (520388526).
    }
}

\author{
    Christian Jobelius Schulz\thanks{Faculty of Economics and Business Administration, University of Duisburg-Essen, Universitätsstraße 12, D--45117 Essen, Germany.}
    \and Christoph Hanck$^{\dagger}$
    \and Simon Hirsch$^{\dagger}$\thanks{Statkraft Trading GmbH, Derendorfer Allee 2a, D--40476, Düsseldorf, Germany.}
    \and Florian Ziel$^{\dagger}$
}
\date{\today}
\maketitle

\begin{abstract}
	\noindent 
	
 We propose an online multivariate distributional forecasting model in which we condition the parameters of a vine copula on covariates. An online algorithm updates the parameters sequentially using only the most recent observations. Existing approaches require full batch re-estimation using all available data at each update step, which makes them computationally slow in large-scale forecasting applications and less suitable for settings in which dependence structures change over time. Simulation results show that the online procedure achieves forecasting performance comparable to that of the batch estimator, while reducing computation time substantially. We illustrate the method in a forecasting application to electricity net-demand across five regions of Great Britain and show that (i) the conditional vine copula model outperforms unconditional Gaussian and vine copula benchmarks and (ii) that online estimation improves forecast accuracy when batch re-estimation is not available. The algorithm is implemented in the \texttt{Python} package \texttt{ondil}.\\
	
\end{abstract}

\section{Motivation}\label{sec:introduction}

Multivariate probabilistic forecasting is central to many problems in economics and the natural sciences. It matters wherever decisions depend not only on the expected level of several related quantities, but also on how these jointly vary. Regional electricity net-demand \citep{gioia2025additive}, asset returns \citep{aquilina2022quantifying}, and weather variables \citep{jobst2024gradient} are prominent applications. These multivariate developments are preceded by a broader shift from point to distributional forecasting \citep[e.g.][]{gneiting2014probabilistic,kneib2023rage,klein2024distributional}. A typical example of such models is the generalized additive model for location, scale, and shape \citep[GAMLSS;][]{rigby2005generalized}, which lets every parameter of a univariate response distribution depend on covariates. Extensions to the multivariate case do the same for the parameters of a multivariate Gaussian distribution \citep[e.g.][]{gioia2025additive,kock2025truly,muschinski2022cholesky}. \citet{hirsch2025online} additionally considers a multivariate $t$-distribution. These approaches inherit the convenient, fully parametric structure of the Gaussian (or $t$) distribution, but also remain tied to a restrictive class of dependence structures.

One way to allow for a richer, non-Gaussian dependence structure is to model the joint distribution via vine copulas. Vine copulas decompose the dependence structure into a sequence of bivariate copula building blocks \citep{vatter2015generalized,vatter2018generalized}. Letting the resulting dependence parameters vary with covariates has a long tradition in this literature, going back to the seminal time-varying copula model of \citet{patton2006modelling}. More recently, generalized additive \citep{vatter2018generalized} and gradient-boosted \citep{jobst2024gradient} pair-copula constructions extend this idea to flexible, nonlinear covariate effects within a vine copula. There, covariates can affect the entire dependence structure, including both its strength, through the dependence parameter, and its shape, through the choice of the bivariate copula family.

At the same time, the data such models are applied to have changed. Decisions in energy and financial markets increasingly rely on data that arrive continuously rather than in periodically released batches. For example, in energy systems, the number of smart electricity meters installed worldwide passed one billion in 2022, a ten-fold increase since 2010, with the number of connected grid sensors and devices projected to roughly double again by 2030 \citep{iea2023unleashing}. Financial markets illustrate the same shift at an even finer time scale. \citet{aquilina2022quantifying} document that latency-arbitrage races between high-frequency traders occur roughly once per minute per stock, with the modal race decided within five to ten millionths of a second. This shift toward high-frequency, streaming, i.e., continuously arriving (pseudo-)real-time data is a defining feature of many modern forecasting applications more broadly \citep{gama2014survey}. It renders repeated full-batch re-estimation of a dependence model, using all historical data every time new observations arrive, impractical. 

This motivates online learning, where model parameters update sequentially with every new observation rather than being re-estimated from scratch. The idea predates the recent surge of interest in streaming data by some time. It builds on the stochastic approximation method of \citet{robbins1951stochastic} and the framework of online convex optimization in machine learning \citep{littlestone1989line,shalev2012online}. More recently, online estimation has also been taken up within macroeconometrics, for instance for the online estimation of dynamic stochastic general equilibrium models \citep{cai2021online} and for online learning in semiparametric econometric models \citep{chen2026online}.

Closer to our setting, \citet{messner2019online} propose to update regularized regression coefficients online by rerunning cyclic coordinate descent algorithms at each update step, using the previous estimates as starting values. Since the coefficients are expected to evolve only gradually over time, the updated estimates remain close to the previous ones, so that the algorithm typically converges after only a few iterations. This idea draws on recursive, $\ell_1$-penalized estimation techniques from the signal processing literature \citep{angelosante2010online}, which develops online coordinate descent (OCD) algorithms for such problems. \citet{hirsch2024online} and \citet{hirsch2025online} employ this machinery to enable online distributional (GAMLSS-type) regression, for univariate and multivariate responses, respectively. 

This article makes two contributions. First, methodologically, we combine distributional forecasting using vine copulas and online learning, and propose an online, regularized generalized linear modeling framework for (conditional) vine copulas. In this framework, the dependence parameter of each pair-copula is linked to covariates through a generalized linear model \citep[GLM;][]{nelder1972generalized} in Kendall's $\tau$, as in \citet{jobst2024gradient}, and updated sequentially via online coordinate descent with ridge-type regularization \citep{friedman2007pathwise,friedman2010regularization}. This allows for fast and stable updates even in high-dimensional covariate settings. To our knowledge, this is the first online estimation procedure for conditional vine copula models. In a simulation study, we show that this online estimator matches the forecasting accuracy of a repeated batch estimator for all but the strongest forgetting factors, while reducing computation time by a factor of up to 10. This factor is even increasing in the overall number of observations, since one update from, say, $1,000$ to $1,001$ observations is as fast as one from $10,000$ to $10,001$ but re-estimation from scratch becomes more expensive with more data.

Second, empirically, we apply our model to the setting of \citet{gioia2025additive}, who forecast the joint distribution of regional electricity net-demand one-step-ahead in Great Britain employing a multivariate Gaussian generalized additive model (GAM) whose covariance structure depends on covariates. Using half-hourly data in a test set covering the entire year 2018, they only update their parameter estimates twelve times, at the beginning of each new month, along the test set. We revisit this setting replacing their Gaussian dependence model with an online, covariate-dependent vine copula model. This allows us to update the parameters also within months in approx. two seconds per update step in our most complex specification. We show that the conditional vine copula model outperforms two unconditional benchmarks, a Gaussian and vine copula model, in our forecasting exercise and that the proposed online procedure improves forecasting performance relative to the forecasting scheme (twelve parameter updates over the test set) entertained by \citet{gioia2025additive}. The model confidence set of \citet{hansen2011model} contains only our conditional vine copula model estimated online with monthly batch re-estimation. This holds when comparing it to unconditional Gaussian and vine copula benchmarks across several forecasting schemes. More generally, forecasting applications of covariate-dependent vine copulas are scarce. In fact, we are only aware of \citet{jobst2024gradient}. Additionally, large-scale, high-dimensional multivariate probabilistic forecasting more broadly remains rare as well, where \citet{hirsch2025online}, \citet{muschinski2022cholesky}, and \citet{gioia2025additive} are, to our knowledge, the few existing examples. 

The remainder of the paper is organized as follows. Section~\ref{sec:methodology} introduces the proposed online copula modeling framework. Section~\ref{sec:simulations} presents a simulation study covering both batch estimation and one-step-ahead forecasting. Section~\ref{sec:applications} applies our approach to jointly forecast regional electricity net-demand in Great Britain. Section~\ref{sec:conclusion} concludes. The Appendix collects the full derivations of the log-likelihood estimator and more details on the simulations and the application. A replication package reproducing all tables and figures is available at \url{https://github.com/chrischu12/online-vine-copulas}, and the estimator is implemented in the open source \texttt{Python} package \texttt{ondil}.

\section{Methodology}\label{sec:methodology}
We propose an online multivariate distributional forecasting model based on vine copulas in which the dependence parameters are allowed to change with covariates in the spirit of \citet{jobst2024gradient}. Our approach separates the modeling of the marginal distributions from the dependence structure. We suppress dependence of the copula parameters on covariates for now. Let
\[
\bm{Y} = (Y_1,\dots,Y_d)^\top
\]
denote a $d$-dimensional random vector with joint distribution $F$ and marginal distributions $F_1,\dots,F_d$. \citet{sklar1959fonctions} shows that the joint density of $\bm{Y}$ decomposes as
$$
	f(y_1,\dots,y_d)
	=
	c\!\left(F_1(y_1),\dots,F_d(y_d)\right)
	\prod_{j=1}^d f_j(y_j),
$$
where $f_j$ denotes the marginal density of $Y_j$ and $c$ the copula density. Applying the probability integral transform to each component yields $U_j = F_j(Y_j) \sim \mathcal{U}(0,1)$, and $\bm{U} = (U_1,\dots,U_d)^\top \in [0,1]^d$ follows a copula $C$ capturing the entire dependence structure of $\bm{Y}$.

\subsection{Model}
\subsubsection{(Conditional) Vine Copulas}
To allow for flexible dependence modeling, \citet{bedford2001probability,bedford2002vines} propose to decompose copula density $c$ into $d(d-1)/2$ bivariate copula densities. Such a decomposition is not unique and can be organized via a regular vine (R-vine) which is a nested sequence of $d-1$ trees $T_\ell = (N_\ell, E_\ell)$, $\ell=1,\dots,d-1$, where the edge set $E_\ell$ is promoted to the node set $N_{\ell+1}$ of the next tree. Let $E = \bigcup_{\ell=1}^{d-1} E_\ell$ denote the set of all $d(d-1)/2$ edges of the vine. Each edge $e \in E$ is associated with a bivariate pair-copula density $c_{a_e,b_e;D_e}$, where $a_e, b_e$ are the conditioned variables and $D_e$ the conditioning set. The vine copula density is then the product over all pair-copula densities, organized tree by tree over the edge sets $E_\ell$,
\begin{align}
	c(\bm{u})
	=
	\prod_{\ell=1}^{d-1}\;\prod_{e \in E_\ell}
	c_{a_e,b_e;D_e}\!\left(u_{a_e \mid D_e},\, u_{b_e \mid D_e};\, \bm{u}_{D_e}\right),
	\label{eq:rvine}
\end{align}
where $\bm{u} = (u_1,\dots,u_d)^\top \in [0,1]^d$ is a realization of $\bm{U}$, and $u_{k \mid D_e} := C_{k \mid D_e}(u_k \mid \bm{u}_{D_e})$ denotes the conditional distribution of $U_k \mid \bm{U}_{D_e}$. In $T_1$, each node represents one of the $d$ variables and each edge models a bivariate pair-copula. In each subsequent tree $T_\ell$, an edge $(a_e, b_e; D_e)$ encodes the dependence between $U_{a_e}$ and $U_{b_e}$ conditioned on $\bm{U}_{D_e}$.

It is common practice to invoke the ``simplifying assumption'' \citep{stoeber2013simplified}, which stipulates that each pair-copula $c_{a_e,b_e;D_e}$ does not depend on the realized value of the conditioning vector $\bm{u}_{D_e}$ for all edges $e$ in the vine. Under this assumption, \eqref{eq:rvine} simplifies to
\begin{align}
	c(\bm{u})
	=
	\prod_{\ell=1}^{d-1}\;\prod_{e \in E_\ell}
	c_{a_e,b_e;D_e}\!\left(u_{a_e \mid D_e},\, u_{b_e \mid D_e}\right).
	\label{eq:rvine_simplified}
\end{align}

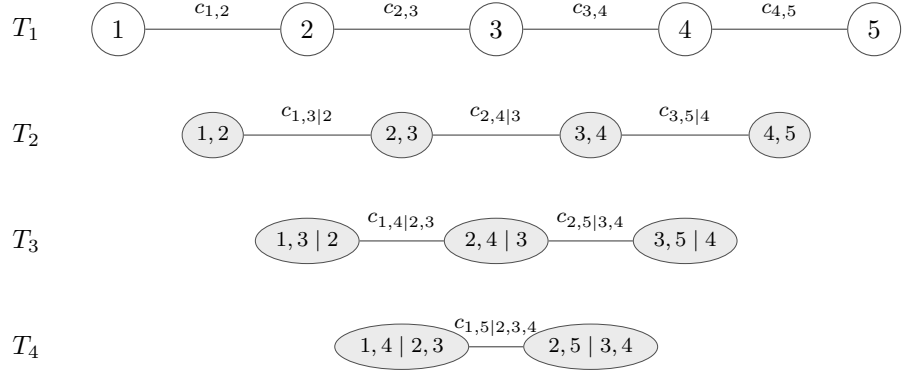
\begin{figure}[t!]
\centering
\tikzset{
  n/.style={circle, draw=black!70, line width=0.4pt, minimum size=7mm, inner sep=0pt, font=\small},
  e/.style={ellipse, draw=black!70, line width=0.3pt, fill=black!8, inner sep=2pt, minimum height=6mm, font=\scriptsize},
  lab/.style={font=\scriptsize},
  tree/.style={font=\itshape\small, anchor=east},
  link/.style={draw=black!60, line width=0.5pt}
}
\begin{tikzpicture}[every node/.style={font=\small}]
  \node[tree] at (-0.3,0) {$T_1$};
  \node[n] (a1) at (0.6,0)   {$1$};
  \node[n] (a2) at (3.1,0)   {$2$};
  \node[n] (a3) at (5.6,0)   {$3$};
  \node[n] (a4) at (8.1,0)   {$4$};
  \node[n] (a5) at (10.6,0)  {$5$};
  \draw[link] (a1)--(a2) node[midway, above, lab] {$c_{1,2}$};
  \draw[link] (a2)--(a3) node[midway, above, lab] {$c_{2,3}$};
  \draw[link] (a3)--(a4) node[midway, above, lab] {$c_{3,4}$};
  \draw[link] (a4)--(a5) node[midway, above, lab] {$c_{4,5}$};
  \node[tree] at (-0.3,-1.4) {$T_2$};
  \node[e] (b1) at (1.85,-1.4)  {$1,2$};
  \node[e] (b2) at (4.35,-1.4)  {$2,3$};
  \node[e] (b3) at (6.85,-1.4)  {$3,4$};
  \node[e] (b4) at (9.35,-1.4)  {$4,5$};
  \draw[link] (b1)--(b2) node[midway, above, lab] {$c_{1,3\mid 2}$};
  \draw[link] (b2)--(b3) node[midway, above, lab] {$c_{2,4\mid 3}$};
  \draw[link] (b3)--(b4) node[midway, above, lab] {$c_{3,5\mid 4}$};
  \node[tree] at (-0.3,-2.8) {$T_3$};
  \node[e] (c1) at (3.10,-2.8) {$1,3\mid 2$};
  \node[e] (c2) at (5.60,-2.8) {$2,4\mid 3$};
  \node[e] (c3) at (8.10,-2.8) {$3,5\mid 4$};
  \draw[link] (c1)--(c2) node[midway, above, lab] {$c_{1,4\mid 2,3}$};
  \draw[link] (c2)--(c3) node[midway, above, lab] {$c_{2,5\mid 3,4}$};
  \node[tree] at (-0.3,-4.2) {$T_4$};
  \node[e] (d1) at (4.35,-4.2) {$1,4\mid 2,3$};
  \node[e] (d2) at (6.85,-4.2) {$2,5\mid 3,4$};
  \draw[link] (d1)--(d2) node[midway, above, lab] {$c_{1,5\mid 2,3,4}$};
\end{tikzpicture}
\caption{D-vine tree structure for $d=5$ as used in our application (see Section~\ref{sec:applications}). }
\label{fig:dvine_general}
\end{figure}

While our framework accommodates general R-vines, we focus on drawable vines (D-vines). D-vines impose a path structure on each tree and are particularly suitable for settings with a natural variable ordering, such as the spatial arrangement of regions of Great Britain in our application. Figure~\ref{fig:dvine_general} illustrates the D-vine tree structure for $d=5$. As an example, the D-vine copula density at $\bm{u}=(u_1,\ldots,u_5)^\top$ may take the form
\begin{align*}
	c(u_1,\ldots,u_5)
	&= c_{1,5;2,3,4}\!\left(u_{1\mid 2,3,4},\,u_{5\mid 2,3,4}\right) \nonumber\\
	&\quad\cdot\,c_{1,4;2,3}\!\left(u_{1\mid 2,3},\,u_{4\mid 2,3}\right)
		\cdot c_{2,5;3,4}\!\left(u_{2\mid 3,4},\,u_{5\mid 3,4}\right) \nonumber\\
	&\quad\cdot\,c_{1,3;2}\!\left(u_{1\mid 2},\,u_{3\mid 2}\right)
		\cdot c_{2,4;3}\!\left(u_{2\mid 3},\,u_{4\mid 3}\right)
		\cdot c_{3,5;4}\!\left(u_{3\mid 4},\,u_{5\mid 4}\right) \nonumber\\
	&\quad\cdot\,c_{1,2}(u_1,u_2)
		\cdot c_{2,3}(u_2,u_3)
		\cdot c_{3,4}(u_3,u_4)
		\cdot c_{4,5}(u_4,u_5),
\end{align*}
with conditional arguments constructed recursively via the $h$-function of each pair-copula,
\[
h(u,v;\theta) := \frac{\partial C(u,v;\theta)}{\partial v}.
\]
Specifically, $u_{j\mid k} = h(u_j,u_k;\theta_{j,k})$ for each $T_1$ edge. Higher-order arguments follow by applying $h$ to pseudo-observations from previous trees. For example,
\[
u_{1\mid 2,3} = h\!\left(u_{1\mid 2},\,u_{3\mid 2};\theta_{1,3\mid 2}\right),
\]
and analogously for the remaining arguments, so that arguments of $T_\ell$ pair-copulas are pseudo-observations produced by the pair-copulas of trees $T_1,\dots,T_{\ell-1}$ \citep{joe1996families}.

Finally, we allow the parameters of each pair-copula to depend on a vector of covariates $\vx$, i.e., $\theta_{a_e,b_e;D_e}(\vx, \vbeta) \in \Theta_{a_e,b_e;D_e}$ turns \eqref{eq:rvine_simplified} into a conditional vine copula density,
\begin{align}
	c(\bm{u};\vx)
	=
	\prod_{\ell=1}^{d-1}\;\prod_{e \in E_\ell}
	c_{a_e,b_e;D_e}\!\left(u_{a_e \mid D_e},\, u_{b_e \mid D_e};\, \theta_{a_e,b_e;D_e}(\vx)\right).
	\label{eq:cond_vine}
\end{align}

\subsubsection{Bivariate Copulas}\label{sec:bivariate_copulas}

The building blocks of the conditional vine copula in \eqref{eq:cond_vine} are the individual pair-copulas, each parameterized by a covariate-dependent parameter $\theta_{a_e,b_e;D_e}(\vx)$. 
In the spirit of \citet{jobst2024gradient}, we model each $\theta_{a_e,b_e;D_e}(\vx)$ through Kendall's $\tau$ via a GLM \citep{nelder1972generalized} as follows. Define
\[
\theta(\vx, \vbeta) := f\!\left(g\big(\eta(\vx)\big)\right),
\]
with linear predictor
\[
\eta := \vx^\top\vbeta,
\qquad \vx \in \SR^p,\quad \vbeta\in\SR^{p},
\]
and Kendall's $\tau$ given by
\[
\tau(\vx,\vbeta) := g\big(\eta(\vx)\big).
\]

To ensure that the restrictions on Kendall's $\tau$ implied by each copula family are respected, we use the inverse link $g:\SR\to(-1,1)$, $g(\eta):=\tanh(\eta/2)$, to map the linear predictor $\eta$ to the range of Kendall's $\tau$, where we assume the link to be correctly specified. A bijective transformation $f:(-1,1)\to\Theta$ then maps Kendall's $\tau\in(-1,1)$ to the copula parameter $\theta\in\Theta\subseteq\SR$; $f$ is available in closed form for all standard parametric families (see Table~\ref{tab:copula_families}). This provides a unified parameterization across copula families (Gaussian, Clayton, Gumbel) with different native parameter spaces. 

In the following, we drop the edge-specific subscripts, $a_e,b_e;D_e$, for notational convenience and write the estimator of $\vbeta$ in matrix notation. Given a sample $\{\vx_i\}_{i=1}^N$, let $\mX\in\SR^{N\times p}$ collect the covariate vectors row-wise and define
\[
\veta := \mX\vbeta \in \SR^N,
\qquad
\vtau := g(\veta) \in (-1,1)^N,
\qquad
\vtheta := f(\vtau) \in \Theta^N,
\]
where $g$ and $f$ are applied element-wise. 

Modelling the dependence parameter through a GLM, we use the Gaussian copula for symmetric dependence, and the Clayton and Gumbel copulas from the Archimedean class for lower and upper tail dependence, respectively. In their original form, the Clayton and Gumbel copulas are only defined for $\tau \geq 0$. We denote this base form by $c_{\mathrm{Family}}$ with $\mathrm{Family}\in\{\mathrm{Clayton},\mathrm{Gumbel}\}$, and extend each to the full range $\tau\in(-1,1)$ by switching to its $90^\circ$ counter-clockwise rotation whenever $\tau<0$,
\begin{align}
c_{\mathrm{Family\,I}}(u_1,u_2;\tau) :=
\begin{cases}
c_{\mathrm{Family}}(u_1,u_2;\tau), & \tau \geq 0, \\
c_{\mathrm{Family}}(u_2,1-u_1;-\tau), & \tau < 0,
\end{cases}
\qquad \mathrm{Family}\in\{\mathrm{Clayton},\mathrm{Gumbel}\}.
\label{eq:family_rotation_i}
\end{align}
A single $\mathrm{Family\,I}$ copula is thus one model in which the sign of $\tau$ selects the rotation. The base copula ($0^\circ$) is selected for $\tau\geq0$ and its $90^\circ$ rotation for $\tau<0$. We obtain a richer set of tail shapes by additionally considering the corresponding survival copula, i.e.,\ the $180^\circ$ rotation of $\mathrm{Family\,I}$,
\begin{align}
c_{\mathrm{Family\,II}}(u_1,u_2;\tau) := c_{\mathrm{Family\,I}}(1-u_1,\,1-u_2;\tau),
\qquad \mathrm{Family}\in\{\mathrm{Clayton},\mathrm{Gumbel}\},
\label{eq:family_rotation_ii}
\end{align}
which by the same sign mechanism is the $180^\circ$ rotation for $\tau\geq0$ and the $270^\circ$ rotation for $\tau<0$. Following \citet{jobst2024gradient}, we therefore model five candidate families (Gaussian, Clayton~I, Clayton~II, Gumbel~I, Gumbel~II) but realize nine distinct copula shapes through the rotations.

\begin{table}[t!]
\centering
\caption{Bijective transformations $f:(-1,1)\to\Theta$ for the considered copula families \citep{jobst2024gradient}.}
\label{tab:copula_families}
\begin{tabular}{lccc}
\toprule
Family & Gaussian & Clayton & Gumbel \\
\midrule
$f(\tau)$ & $\sin\!\left(\dfrac{\pi}{2}\,\tau\right)$ & $\dfrac{2\tau}{1-|\tau|}$ & $\dfrac{\operatorname{sign}(\tau)}{1-|\tau|}$ \\[8pt]
\bottomrule
\end{tabular}
\end{table}

\subsection{Estimation}

We follow the inference function for margins (IFM) approach \citep{joe1996estimation}, which decouples the estimation of marginal distributions from the dependence structure. First, each marginal $F_j$, potentially dependent on covariates, $\vx$, is estimated separately. Second, the observations are transformed to pseudo-uniform values $\hat{u}_{j,i} = \hat{F}_j(y_{j,i})$, on which the vine copula is estimated. This two-step procedure avoids joint optimization over all marginal and copula parameters simultaneously. Simulation studies in \citet{joe1996estimation} and \citet{joe2005asymptotic} show that the IFM method retains high asymptotic efficiency compared to full maximum likelihood estimation. We now discuss the second step of estimating the vine copula parameters. 

\subsubsection{(Conditional) Vine Copulas}\label{sec:vine_estimation}

As mentioned above, the vine copula density in \eqref{eq:cond_vine} is a product over $d(d-1)/2$ pair-copula densities. Throughout this paper we assume the vine structure to be known in advance and fix a D-vine. This implies that estimating the vine copula boils down to estimating five pair-copula parameters (one per copula family) per edge in the vine and selecting the best fitting family among the five candidates using the Akaike Information Criterion (AIC). As is standard in the vine copula literature, we adopt a sequential top-down procedure \citep{aas2009pair}, where pair-copulas are estimated tree by tree, starting from $T_1$, with all edges within each tree estimated independently. This avoids joint optimization over all $d(d-1)/2$ pair-copula parameters. The same estimation and family-selection procedure as in $T_1$ is then applied to each remaining tree $T_\ell$, $\ell=2,\dots,d-1$. 
 When the vine structure is unknown, it can be selected via a maximum spanning tree algorithm on pairwise empirical Kendall's $\tau$ values \citep{dissmann2013selecting}. This poses no obstacle to the proposed estimation procedure and is also implemented in the \texttt{Python} package \texttt{ondil}. Recent advances in vine structure selection \citep{vatter2025throwing} could be incorporated as well. We leave this for future research.

\subsubsection{Bivariate Copulas}\label{sec:bivariate_estimation}

We now describe the estimation of a single pair-copula, which is the step carried out for each of the edges in the vine copula. This amounts to maximizing a copula log-likelihood per edge. For $\vu\in[0,1]^2$ and copula parameter $\theta\in\Theta$, define
\[
\ell(\vu,\vx;\vbeta) = \log c\!\left(\vu;\theta(\vx,\vbeta)\right),
\qquad \theta(\vx,\vbeta) = f\!\left(g(\vx^\top\vbeta)\right).
\]

We propose a penalized maximum likelihood (ridge) estimator for $\vbeta$ to ensure stable estimation and prevent overfitting, especially in high-dimensional settings. Given a sample $\{(\vu_i,\vx_i)\}_{i=1}^N$, the penalized log-likelihood reads
\[
\ell_c(\vbeta) = \ell(\vbeta) - \frac{1}{2}\lambda\,\vbeta^\top\vbeta,
\qquad
\ell(\vbeta)=\sum_{i=1}^N \ell(\vu_i,\vx_i;\vbeta),
\]
and the penalized maximum likelihood estimator is
\[
\widehat{\vbeta}_\lambda = \arg\max_{\vbeta\in\SR^p} \ell_c(\vbeta).
\]
We recast this problem as an iteratively reweighted ridge regression \citep{vatter2015generalized} and defer details to Appendix~\ref{app:derivations_newton_raphson}. Eventually, we obtain a Newton--Raphson update step for the regression coefficients at optimization iteration $t+1$,
\begin{equation}
  \label{eq:newton_raphson}
\widehat{\vbeta}^{[t+1]}
=
\Bigl(\mX^\top \mW^{[t]}\mX + \lambda\,\mI\Bigr)^{-1}
\mX^\top \mW^{[t]} \vz^{[t]},
\end{equation}
with weights, $\mW$, linear predictor, $\veta$, working vector, $\vz$, and score, $\vs$ defined as
\begin{align*}
  \mW^{[t]}   &= -\operatorname{diag}\!\left(\frac{\partial^2 \ell(\vbeta)}{\partial \veta\,\partial \veta^\top}\right)^{\![t]}, \\
	\veta^{[t]} &= \mX\widehat{\vbeta}^{[t]}, \\
	\vz^{[t]}   &= \veta^{[t]} + \bigl(\mW^{[t]}\bigr)^{-1}\vs^{[t]}, \\
	\vs^{[t]}   &= \frac{\partial \ell(\vbeta)}{\partial \veta}\bigg|^{[t]}.
\end{align*}

In each update step $t$, we use a cyclic coordinate descent algorithm \citep{friedman2007pathwise, friedman2010regularization} to estimate $\vbeta$. This allows us to efficiently compute the entire regularization path for a grid of $\lambda$ values, which is necessary for selecting the optimal $\lambda$ via an information criterion.
We do this by using the effective degrees of freedom (EDF) of the penalized estimator \citep{vatter2018generalized,umlauf2018bamlss,groll2019lasso}
\begin{equation*}
	\operatorname{edf}(\lambda)
	:= \operatorname{tr}\!\left[
	\mX^\top \mW \mX
	\left(\mX^\top \mW \mX + \lambda\,\mI\right)^{-1}
	\right].
\end{equation*}

With the $\operatorname{edf}(\lambda)$ at hand, information criteria take their usual form,
\begin{equation}
	\label{eq:ic}
	\operatorname{AIC}(\lambda) = -2\,\ell(\widehat{\vbeta}_\lambda)
	+ 2\,\operatorname{edf}(\lambda),
	\qquad
	\operatorname{BIC}(\lambda) = -2\,\ell(\widehat{\vbeta}_\lambda)
	+ \log(n)\,\operatorname{edf}(\lambda),
\end{equation}
and the optimal $\lambda^\ast$ is selected as the minimizer along the grid used in the coordinate descent path. 

Appendix \ref{app:derivations_newton_raphson} provides the full derivation of the score and Hessian for the copula log-likelihood, which are used to compute $\vs$ and $\mW$, respectively, in each update step. To give a concrete idea, consider the Gaussian copula, whose parameter we denote by $\theta$. The score $\vs$ is the derivative of the log-density with respect to the linear predictor $\eta$, which we obtain by the chain rule along $\eta\mapsto\tau\mapsto\theta$,
\[
\frac{\partial \ell}{\partial \eta}
= \underbrace{\frac{\partial \ell}{\partial \theta}}_{\text{log-density}}
\cdot \underbrace{f'(\tau)}_{\text{bijection } \theta=\sin(\frac{\pi}{2}\tau)}
\cdot \underbrace{g'(\eta)}_{\text{link } \tau=\tanh(\eta/2)} ,
\]
so we need three derivatives: the bijection $f'(\tau)$, the link $g'(\eta)$, and the log-density derivative. The latter is
\[
\frac{\partial \ell}{\partial \theta}
= \frac{\theta(1-\theta^2) - \theta\bigl(\Phi^{-1}(u_1)^2+\Phi^{-1}(u_2)^2\bigr) + (1+\theta^2)\Phi^{-1}(u_1)\Phi^{-1}(u_2)}{(1-\theta^2)^2},
\]
where $\Phi$ denotes the standard normal CDF. The weight $\mW$ requires the corresponding second derivative $\partial^2\ell/\partial\eta^2$, which adds $g''$, $f''$, and $\partial^2\ell/\partial\theta^2$ by the same logic. The Gumbel and Clayton families follow an identical pattern, differing only in $f$ and $\ell$. See Appendix~\ref{app:normal_copula} for the full derivation of the Gaussian copula score and Hessian.

\begin{algorithm}[t!]
\caption{Newton--Raphson with ridge penalty (batch)}
\label{alg:irls}
\DontPrintSemicolon
\KwIn{Copula $c$, data $\{(\vu_i,\vx_i)\}_{i=1}^N$, initial $\widehat{\vbeta}^{[0]}$, penalty grid $\Lambda$}
\KwOut{$\widehat{\vbeta}$, $\widehat{\veta}$, $\widehat{\vtau}$, $\widehat{\vtheta}$}

\For{$t = 0,1,2,\dots$ }{

  Compute $\veta^{[t]} = \mX\widehat{\vbeta}^{[t]}$ and $\vtau^{[t]} = g(\veta^{[t]})$\;

  Compute score $\vs^{[t]} = \dfrac{\partial \ell(\vbeta)}{\partial \veta}\bigg|^{[t]}$ and weight matrix $\mW^{[t]} = -\operatorname{diag}\!\left(\dfrac{\partial^2 \ell(\vbeta)}{\partial \veta\,\partial \veta^\top}\right)^{\![t]}$\;

  Form working vector $\vz^{[t]} = \veta^{[t]} + \bigl(\mW^{[t]}\bigr)^{-1}\vs^{[t]}$\;

  Form statistics $\mG^{[t]} = \mX^\top\mW^{[t]}\mX$ and $\vh^{[t]} = \mX^\top\mW^{[t]}\vz^{[t]}$\;

  \For{$\lambda \in \Lambda$}{
    Solve $\bigl(\mG^{[t]} + \lambda\,\mI\bigr)\,\widehat{\vbeta}^{[t+1]}_\lambda = \vh^{[t]}$ by cyclic coordinate descent, warm-started across $\Lambda$\;
  }

  Select $\lambda^\ast = \operatorname{arg\,min}_{\lambda\in\Lambda}\operatorname{AIC}(\lambda)$ and set $\widehat{\vbeta}^{[t+1]} = \widehat{\vbeta}^{[t+1]}_{\lambda^\ast}$\;

  \If{$\|\widehat{\ell}^{[t+1]} - \widehat{\ell}^{[t]}\| < \varepsilon$}{
    \textbf{break}\;
  }
}

\end{algorithm}
We disentangle the estimation procedure in Algorithm~\ref{alg:irls} to clarify the connection to the online procedure in Algorithm~\ref{alg:online_update}.  It is based on the same Newton--Raphson updates but relies on sequentially updated statistics $\mG_{N+1}$ and $\vh_{N+1}$ to be introduced in the following.

\subsubsection{Online Estimation}\label{sec:online_estimation}

The basic idea of the online estimation procedure is to rerun the cyclic coordinate descent algorithm at each online update step (when a new data point arrives), using the coefficient estimates from the previous step as starting values. Since the coefficients are assumed to evolve only gradually over time, the updated estimates remain close to the previous ones, so that the algorithm converges after only a few iterations. This online coordinate descent \citep[OCD;][pp. 3440]{angelosante2010online} algorithm has been developed in the signal-processing literature, where it is applied to lasso estimation of linear regression models. \citet{messner2019online} transfer the idea to vector autoregressive models and \citet{hirsch2024online} and \citet{hirsch2025online} to univariate and multivariate GAMLSS-type distributional regression, respectively. We apply a similar strategy to our iteratively reweighted ridge regression problem by updating statistics that summarize the information contained in the data seen so far.
The main step occurs in the Newton--Raphson step, where we need to compute $\mG = \mX^\top \mW \mX$ and $\vh = \mX^\top \mW \vz$ (see \eqref{eq:newton_raphson}). In the batch setting, these are computed from the full dataset. For the online algorithm, the subscript $N$ denotes the current sample size, so 
\begin{align*}
	\mG_N &= \mX_ N^\top \mW_N \mX_N, \\
	\vh_N &= \mX_ N^\top \mW_N \vz_ N,
\end{align*}
with $\mG_N\in\SR^{p \times p}$ and $\vh_N\in\SR^{p}$ summarizing all information up to and including observation $N$.

\begin{algorithm}[t!]
\caption{Newton--Raphson with ridge penalty (online)}
\label{alg:online_update}
\DontPrintSemicolon
\KwIn{Copula $c$, new observation $(\vu_{N+1},\vx_{N+1})$, current estimates $\widehat{\vbeta}_N$, sufficient statistics $\mG_N$, $\vh_N$, forgetting factor $\gamma$, penalty grid $\Lambda$}
\KwOut{$\widehat{\vbeta}_{N+1}$, $\widehat{\eta}_{N+1}$, $\widehat{\tau}_{N+1}$, $\widehat{\theta}_{N+1}$}

Initialize $\widehat{\vbeta}_{N+1}^{[0]} = \widehat{\vbeta}_N$ 

\For{$t = 0,1,2,\dots$ }{

  Compute $\eta_{N+1}^{[t]} = \vx_{N+1}^\top\widehat{\vbeta}_{N+1}^{[t]}$ and $\tau_{N+1}^{[t]} = g\!\left(\eta_{N+1}^{[t]}\right)$\;

  Compute scalar score $s_{N+1}^{[t]} = \dfrac{\partial \ell_{N+1}}{\partial \eta}\bigg|^{[t]}$ and scalar weight $w_{N+1}^{[t]} = -\dfrac{\partial^2 \ell_{N+1}}{\partial \eta^2}\bigg|^{[t]}$\;

  Form scalar working vector $z_{N+1}^{[t]} = \eta_{N+1}^{[t]} + \bigl(w_{N+1}^{[t]}\bigr)^{-1} s_{N+1}^{[t]}$\;

  Update statistics\;
  $\mG_{N+1}^{[t]} = (1-\gamma)\,\mG_N + w_{N+1}^{[t]}\,\vx_{N+1}\vx_{N+1}^\top$\;
  $\vh_{N+1}^{[t]} = (1-\gamma)\,\vh_N + w_{N+1}^{[t]}\,\vx_{N+1}\,z_{N+1}^{[t]}$\;

  Update $\widehat{\vbeta}_{N+1,\lambda}^{[t+1]}$ for each $\lambda \in \Lambda$ via Online Coordinate Descent\label{line:ocd}\;

  Select $\lambda^\ast = \operatorname{arg\,min}_{\lambda\in\Lambda}\operatorname{AIC}(\lambda)$ and set $\widehat{\vbeta}_{N+1}^{[t+1]} = \widehat{\vbeta}_{N+1,\lambda^\ast}^{[t+1]}$\;

  \If{$|\widehat{\ell}_{N+1}^{[t+1]} - \widehat{\ell}_{N+1}^{[t]}| < \varepsilon$}{
    \textbf{break}\;
  }
}
\end{algorithm}

Upon arrival of the new observation $(z_{N+1},\vx_{N+1})$, $\mG_N$ and $\vh_N$ are updated as
\begin{equation}
\begin{aligned}
    \mG_{N+1}
    &= (1-\gamma)\,\mG_N
    + \,w_{N+1}\,\vx_{N+1}\vx_{N+1}^\top, \\
    \vh_{N+1}
    &= (1-\gamma)\,\vh_N
    + \,w_{N+1}\,\vx_{N+1}\,z_{N+1},
\end{aligned}
\label{eq:gramian_update}
\end{equation}
with $\vx_{N+1}\in\SR^{p}$ the covariate vector of the new observation, scalar weight $w_{N+1}$, and scalar working vector $z_{N+1}$. Thus, $\vx_{N+1}\vx_{N+1}^\top\in\SR^{p\times p}$ and $\vx_{N+1}z_{N+1}\in\SR^{p}$ match $\mG_N$ and $\vh_N$, respectively. Here, we essentially perform a running mean with exponential forgetting. The forgetting factor $\gamma\in[0,1)$ controls how much weight is placed on past information relative to the new observation. $\gamma=0$ recovers the batch update which gives equal weight to all observations, whereas higher values of $\gamma$ downweight older observations more aggressively. There is a trade-off in the choice of $\gamma$. Smaller values lead to more precise estimates but are less responsive to changes in the underlying DGP, while larger values allow for quicker adaptation but may introduce more variability into the estimates.

The coefficient estimates along the penalty path are then updated as described in line~\ref{line:ocd} of Algorithm~\ref{alg:online_update} via OCD, which updates one coefficient at a time while keeping the others fixed. Alternatively, the ridge regression problem could be solved by traditional recursive least squares, which for each $\lambda \in \Lambda$ computes
\begin{equation}
    \widehat{\vbeta}_{N+1,\lambda}^{[t+1]} = \Bigl(\mG_{N+1}^{[t]} + \lambda\,\mI\Bigr)^{-1}\vh_{N+1}^{[t]}.
    \label{eq:ridge_closed_form}
\end{equation}
However, this would incur repeated inversion of the potentially high-dimensional matrix $\bigl(\mG_{N+1}^{[t]} + \lambda\,\mI\bigr)$ for each $\lambda \in \Lambda$, which is both computationally expensive and numerically unstable. The OCD algorithm avoids this inversion entirely. See Appendix~\ref{app:ocd} for details on the OCD update step.

\section{Simulations}\label{sec:simulations}

We conduct two simulation studies. First, in Section~\ref{sec:simulation_batch}, we focus on the estimation of conditional vine copula models in a batch setting. Second, in Section~\ref{sec:online_vs_batch}, we evaluate the forecasting performance of the proposed online estimation procedure compared to a repeated batch approach. 

\subsection{Batch Estimation}\label{sec:simulation_batch}

We consider a conditional D-vine copula model in $d = 5$ dimensions, i.e., we need to estimate $d(d-1)/2 =10$ edges. The data-generating process (DGP) is given by
\begin{align}\label{eq:dgp}
	\tau(\vx) = g\!\left(\vx^\top\vbeta\right) = \tanh\!\left(\tfrac{1}{2}\vx^\top\vbeta\right),
\end{align}
where, as in \citet{vatter2018generalized}, $\vx = (x_1,\dots,x_{10})^\top \in \SR^{10}$ and $x_j \sim \mathrm{Bernoulli}(0.5)$ for $j=1,\dots,5$, $x_j \sim \mathcal{N}(0,1)$ for $j=6,\dots,10$.    We set
\begin{align*}
	\vbeta = \tfrac{1}{4} \big(1,\, 1,\,-1,\, 0,\, 0,\, 1,\,-1,\,-1,\, 0,\, 0\big)^\top,
\end{align*}
so that only six components of $\vx$ have a non-zero effect on $\tau(\vx)$. Each pair-copula is drawn with equal probability from the five candidate families introduced in Section~\ref{sec:bivariate_copulas}: Gaussian, Clayton~I, Clayton~II, Gumbel~I, and Gumbel~II. As in \eqref{eq:family_rotation_i} and \eqref{eq:family_rotation_ii}, the sign of the conditional Kendall's $\tau$ selects the rotation pointwise within each Family~I and Family~II copula, so that the Clayton and Gumbel families jointly span all four rotations $\{0^\circ,90^\circ,180^\circ,270^\circ\}$ and thereby cover the full range $\tau\in(-1,1)$. Model selection is based on the AIC. We use 500 Monte Carlo iterations for sample sizes of $n \in \{1000, \ 5000\}$. We estimate the model using the unregularized version of our approach, i.e., tuning parameter $\lambda$ is set to 0. We do so to allow for unbiased estimates.

\begin{figure}[t!]
    \centering
    \includegraphics[width=\linewidth]{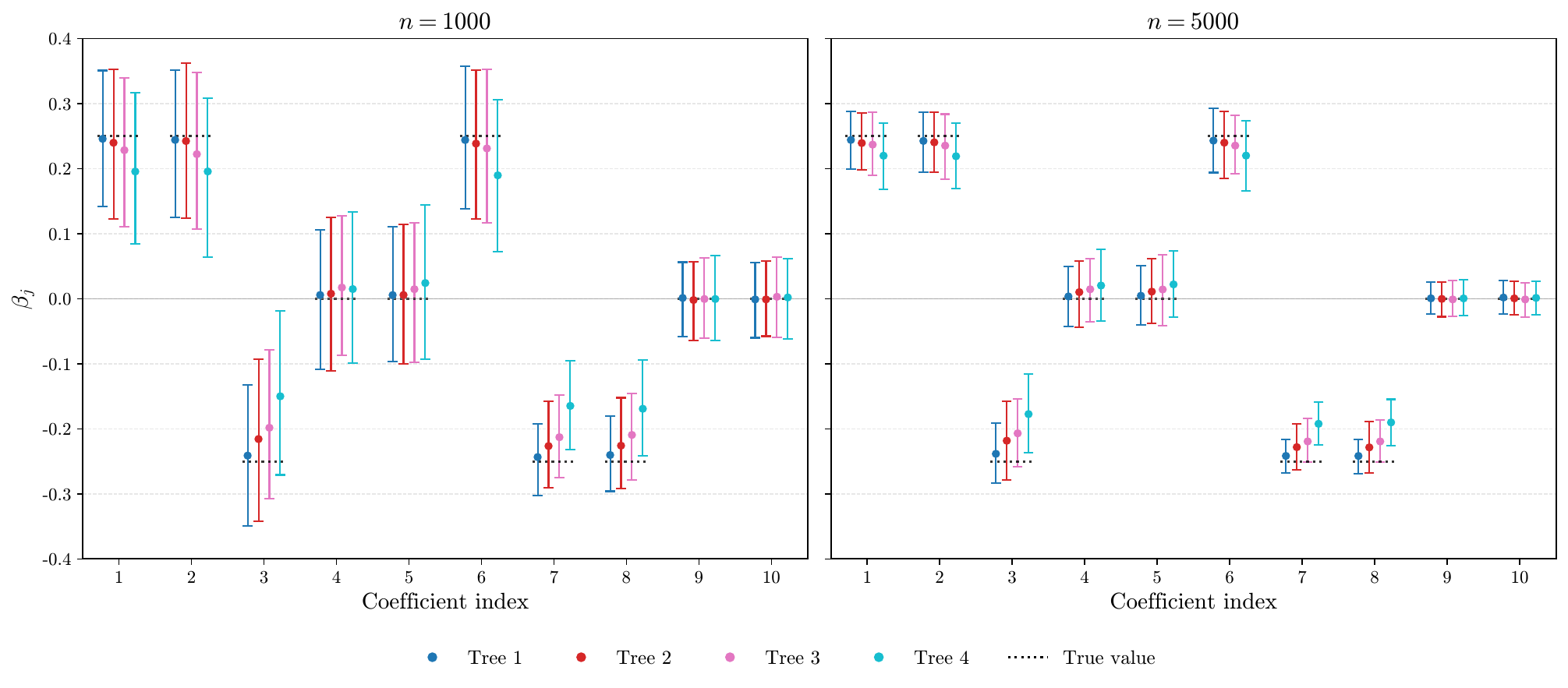}
    \caption{Monte Carlo distribution of $\beta_j$ across trees for $n=1000$ (left panel) and $n=5000$ (right panel) based on the unregularized model with tuning parameter $\lambda = 0$. The true values are indicated by dotted lines.}
    \label{fig:beta_estimates}
\end{figure}

Figure~\ref{fig:beta_estimates} shows the empirical distribution of the estimated $\beta_j$ across trees. We see that the estimates are centered around their true values, with the dispersion shrinking as the sample size increases. This is expected from consistency results for penalized likelihood estimators \citep{green1987penalized}. The results demonstrate that the proposed estimation procedure can successfully recover the underlying parameter values in a batch setting.
Because estimation error in $T_{\ell-1}$ propagates into the pseudo-observations of $T_\ell$, the effective signal of covariates on Kendall's $\tau$ weakens with tree depth. This is a pattern that is documented by \citet{vatter2018generalized} and \citet{jobst2024gradient} as well. In our application in Section~\ref{sec:applications} we therefore restrict covariate effects to the first tree and use intercept-only models from $T_2$ onwards.

\subsection{Online vs Batch Forecasting}\label{sec:online_vs_batch}

We use the same DGP as in the previous Section~\ref{sec:simulation_batch} and split the observations into a training ($n_{\text{train}} = 700$) and a test set ($n_{\text{test}} = 300$). We additionally add noise through adding 10 irrelevant Gaussian regressors to the DGP in \eqref{eq:dgp}. One-step-ahead forecasts are generated over the test period. At each step we estimate the vine copula on all observations up to observation $N$ and predict $N+1$ using the newly observed covariates. We then evaluate its predictive density at the realized observation ($N+1$) to obtain the log score (LS).

\begin{figure}[t!]
	\centering
	\includegraphics[width=\textwidth]{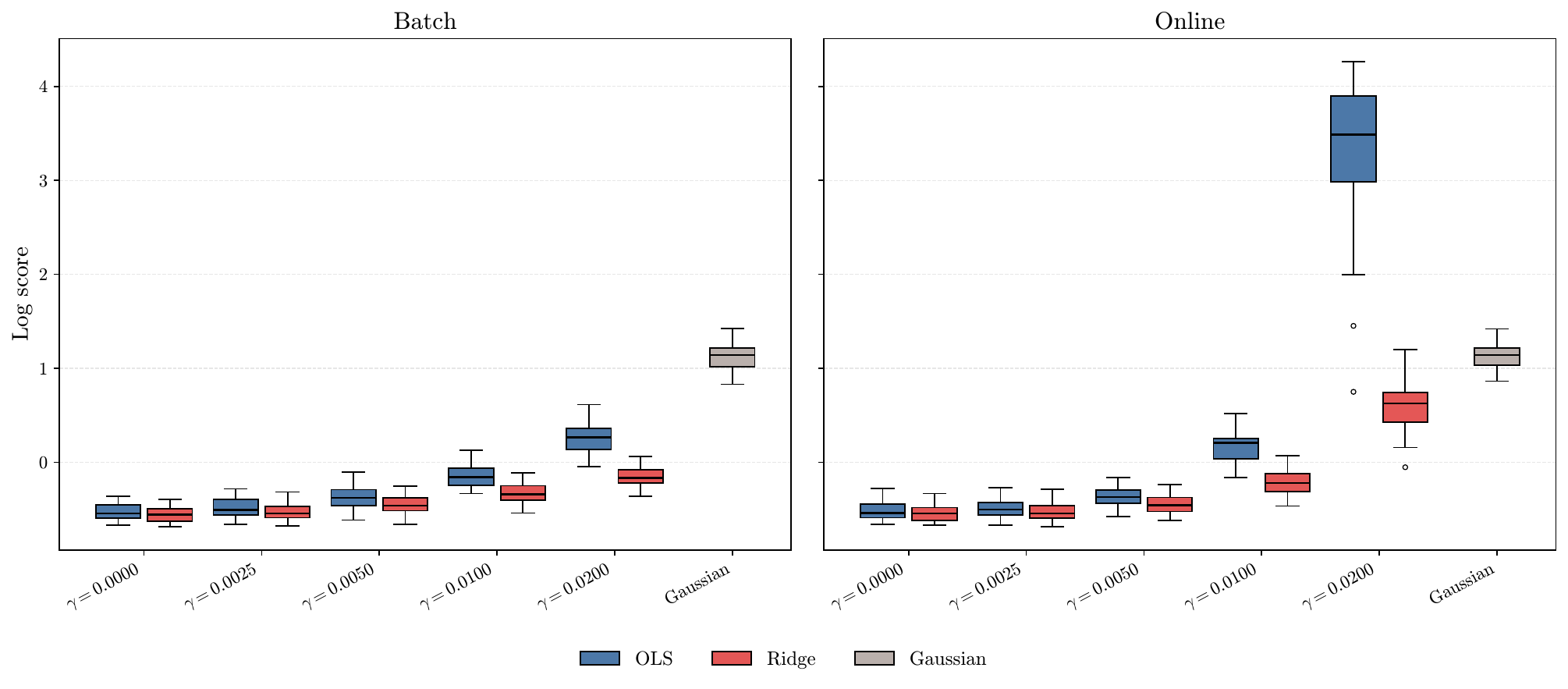}
	\caption{Distributions of log scores across updating schemes (Batch vs.\ Online; left vs.\ right panel), estimation methods (OLS vs.\ Ridge; blue vs.\ red), and forgetting factors $\gamma$ ($x$-axis), based on $50$ Monte Carlo iterations. The online procedure updates the parameters sequentially as each new observation arrives, whereas the batch procedure refits on all available data. The forgetting factor $\gamma$ governs the degree of forgetting in both procedures. $\gamma=0$ corresponds to no forgetting; larger values place more weight on recent observations. The Gaussian specification is based on an unconditional Gaussian copula.}
	\label{fig:ols_vs_ridge_boxplots}
\end{figure}

We consider two different forecasting schemes: an online approach and a repeated batch procedure. The online approach updates the model parameters sequentially as new data become available in the test period (after an initial batch fit on the training data). The batch procedure refits the model on all available data for each new forecasting step. Furthermore, we vary the estimation method and the forgetting factor. Specifically, we consider $\ell_2$ (ridge) vs unpenalized estimation, where the penalty applies to the slopes only and the intercept is left unpenalized, and forgetting factors $\gamma \in \{1/50,\, 1/100,\, 1/200,\, 1/400,\, 0\}$, i.e.,\ $\gamma \in \{0.02,\, 0.01,\, 0.005,\, 0.0025,\, 0\}$, corresponding to rolling window sizes of $50$, $100$, $200$, $400$, and $\infty$, respectively. This is applied for both the batch and the online updating scheme. The results are based on 50 Monte Carlo iterations. Forecasting performance is evaluated using the (negative) LS, which is a proper scoring rule for probabilistic forecasts and particularly suited for comparing copula-based models \citep{fissler2026compare}.

Figure~\ref{fig:ols_vs_ridge_boxplots} illustrates several aspects of the proposed online conditional vine copula model. First, we demonstrate that conditional vine copula specifications outperform unconditional Gaussian copula models in capturing complex dependence structures. The negative LS for the unconditional Gaussian copula is highest across almost all specifications in Figure~\ref{fig:ols_vs_ridge_boxplots}. Second, we show that regularization provides better forecasting performance compared to unregularized estimation, independent of whether estimation is performed in batch or online settings. The negative LS is smaller for all ridge specifications compared to its OLS counterpart. Third, we illustrate, drawing additionally on Table~\ref{tab:time_ridge} that for this particular forecasting exercise, there is essentially no trade-off between computational efficiency and estimation accuracy. While online estimation is considerably faster than repeatedly refitting the batch model, it does not sacrifice estimation accuracy in almost all of our specifications. Only for the two largest forgetting factors does online estimation fall behind its batch counterpart. It does so slightly at $\gamma = 0.01$ and more substantially at $\gamma = 0.02$, i.e., for rolling window sizes of 100 and 50. Computation times for unpenalized estimation can be found in Appendix~\ref{app:computationtimesim}. Fourth, drawing on Figure~\ref{fig:break_vs_stable_gamma} in Appendix~\ref{app:structuralbreak}, when the DGP changes during the test set, mild forgetting, $\gamma = 0.0025$, improves forecasting performance compared to using all available data ($\gamma = 0$). The reason is that forgetting trades sample size against adaptivity. When the dependence structure is stable, only the effective sample size matters, so the LS is minimized without forgetting. Under a structural break in the coefficients, discounting past observations instead lets the estimates adapt to the new regime.
\begin{table}[t!]
    \centering
\caption{Average vine estimation time per update under ridge regularization.}
\label{tab:time_ridge}
\input{tables/table_vine_time_ridge_avg.tex}
	\captionsetup[table]{font=small} 
	\caption*{\textit{Notes:} We compute the average estimation time per update across all observations in the test set in seconds. The batch vine is estimated from scratch after each new observation, while the online vine is updated sequentially. The forgetting factor $\gamma$ controls the window size applied to the vine estimation.}
\end{table}

\section{Forecasting Regional Electricity Net-Demand in Great Britain}\label{sec:applications}

Forecasts of regional electricity net-demand, defined as electricity demand net of
embedded generation, are a key input for reliable and cost-efficient power system
operation and energy trading. Although such forecasts are often produced separately
for each region \citep[see][for a review on distributional load forecasting]{hong2016probabilistic}, many operational tasks, such as managing interregional power flows, and quantifying imbalance risk, require spatially coherent joint forecasts that account for cross-regional dependence. However, the literature on joint probabilistic forecasting of regional net-demand is scarce. In fact, we are only aware of the paper by \citet{gioia2025additive} who model the joint distribution of regional net-demand using a GAM based on gradient boosting with a multivariate Gaussian specification in which the covariance matrix itself depends on covariates.

Conditioning the dependence structure on covariates in this way serves two purposes,
which we retain in our application. First, it makes the model interpretable. Time-varying dependence 
can be directly related to physical weather regimes. Second, conditioning on weather covariates lets the joint forecast exploit this forward-looking information rather than relying solely on
historical dependence patterns.

The Gaussian specification of \citet{gioia2025additive}, however, has two
limitations. First, it only allows covariates to shift the strength of dependence,
not its shape. This matters because simultaneous high net-demand events are
particularly relevant for security of supply. Since the
Gaussian copula has zero tail dependence by construction \citep{joe2014dependence},
it can understate the probability of joint extreme net-demand outcomes. At the same
time, Gaussian dependence may still be adequate in less extreme or more regular
regimes and should remain available as a special case. Second, the parameters of the multivariate normal distribution are re-estimated only once per month in the test set so that intra-month
variation in dependence is not captured.

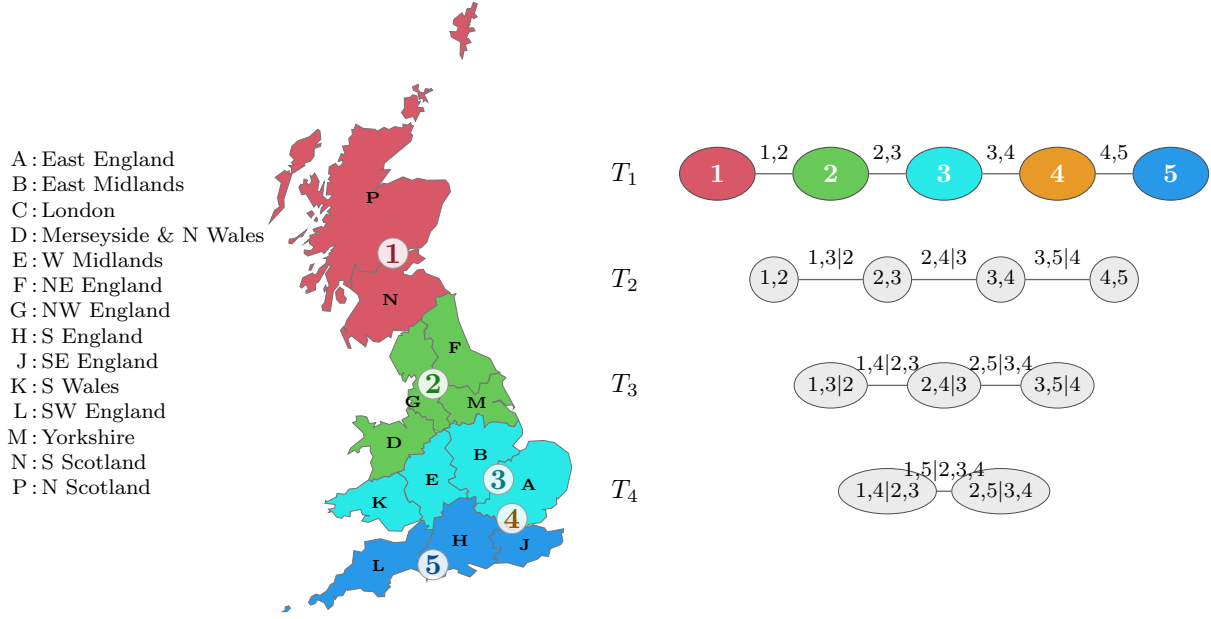
\begin{figure}[t!]
  \centering
  \begin{minipage}[c]{0.20\textwidth}
    \centering
    \scriptsize
    \setlength{\tabcolsep}{2pt}
    \begin{tabular}{@{}r@{\,:\,}l@{}}
      A & East England \\
      B & East Midlands \\
      C & London \\
      D & Merseyside \& N Wales \\
      E & W Midlands \\
      F & NE England \\
      G & NW England \\
      H & S England \\
      J & SE England \\
      K & S Wales \\
      L & SW England \\
      M & Yorkshire \\
      N & S Scotland \\
      P & N Scotland \\
    \end{tabular}
  \end{minipage}\hfill
  \begin{minipage}[c]{0.22\textwidth}
    \centering
    \input{figures/uk_map.tex}
  \end{minipage}\hfill
  \begin{minipage}[c]{0.56\textwidth}
    \centering
    \begin{tikzpicture}[
      every node/.style={font=\small},
      n/.style={ellipse, draw=black!70, line width=0.4pt, minimum width=10mm, minimum height=7mm, inner sep=0pt, text=white, font=\bfseries\small},
      e/.style={ellipse, draw=black!70, line width=0.3pt, fill=black!8, inner sep=1pt, minimum height=6mm, font=\scriptsize},
      lab/.style={font=\scriptsize},
      tree/.style={font=\itshape\small, anchor=east},
      link/.style={draw=black!60, line width=0.5pt}
    ]
      \definecolor{r1}{HTML}{D85868}
      \definecolor{r2}{HTML}{68C858}
      \definecolor{r3}{HTML}{28E8E8}
      \definecolor{r4}{HTML}{E89A28}
      \definecolor{r5}{HTML}{2898E8}

      \node[tree] at (-0.3,0) {$T_1$};
      \node[n, fill=r1] (a1) at (0.6,0)  {1};
      \node[n, fill=r2] (a2) at (2.1,0)  {2};
      \node[n, fill=r3] (a3) at (3.6,0)  {3};
      \node[n, fill=r4] (a4) at (5.1,0)  {4};
      \node[n, fill=r5] (a5) at (6.6,0)  {5};
      \draw[link] (a1)--(a2) node[midway, above, lab] {1,2};
      \draw[link] (a2)--(a3) node[midway, above, lab] {2,3};
      \draw[link] (a3)--(a4) node[midway, above, lab] {3,4};
      \draw[link] (a4)--(a5) node[midway, above, lab] {4,5};

      \node[tree] at (-0.3,-1.4) {$T_2$};
      \node[e] (b1) at (1.35,-1.4) {1,2};
      \node[e] (b2) at (2.85,-1.4) {2,3};
      \node[e] (b3) at (4.35,-1.4) {3,4};
      \node[e] (b4) at (5.85,-1.4) {4,5};
      \draw[link] (b1)--(b2) node[midway, above, lab] {1,3$\mid$2};
      \draw[link] (b2)--(b3) node[midway, above, lab] {2,4$\mid$3};
      \draw[link] (b3)--(b4) node[midway, above, lab] {3,5$\mid$4};

      \node[tree] at (-0.3,-2.8) {$T_3$};
      \node[e] (c1) at (2.1,-2.8) {1,3$\mid$2};
      \node[e] (c2) at (3.6,-2.8) {2,4$\mid$3};
      \node[e] (c3) at (5.1,-2.8) {3,5$\mid$4};
      \draw[link] (c1)--(c2) node[midway, above, lab] {1,4$\mid$2,3};
      \draw[link] (c2)--(c3) node[midway, above, lab] {2,5$\mid$3,4};

      \node[tree] at (-0.3,-4.2) {$T_4$};
      \node[e] (d1) at (2.85,-4.2) {1,4$\mid$2,3};
      \node[e] (d2) at (4.35,-4.2) {2,5$\mid$3,4};
      \draw[link] (d1)--(d2) node[midway, above, lab] {1,5$\mid$2,3,4};
    \end{tikzpicture}
  \end{minipage}
  \caption{The 14 GB GSP areas (left) grouped into five macro regions and the corresponding D-vine structure on the macro regions (right). The macro regions are ordered from north to south: 1~Scotland, 2~Northern, 3~Midlands, 4~London, 5~Southern.}
  \label{fig:uk_map}
\end{figure}

These considerations imply that a suitable model should be: (i) multivariate, so that aggregate and cross-regional risks can be evaluated coherently;
(ii) conditional on interpretable covariates, so that time variation in dependence can
be linked to economically meaningful weather regimes; (iii) flexible enough to allow
for non-Gaussian and tail-dependence while retaining Gaussian dependence as
a potential case; and (iv) estimable online, so that the
dependence structure can adapt as new observations arrive rather than being fixed
within each re-estimation window. Online conditional vine copulas (in combination with a suitable model for the marginals) satisfy these
requirements. They decompose the multivariate dependence structure into bivariate pair
copulas \citep{aas2009pair}, allow pair-specific copula families such as Gaussian,
Gumbel, Clayton, and rotated variants, and let the corresponding dependence
parameters vary with edge-specific weather covariates. Our online estimation
algorithm, in turn, updates these parameters sequentially as new data become
available. We will, thus, investigate in this application whether our proposed conditional vine copula model can improve forecasting performance compared to unconditional Gaussian and vine copula specifications. Additionally, we will analyze whether online estimation can enhance forecasting performance compared to a repeated batch approach, in which the parameters are updated only twelve times in the test set. 

\begin{table}[t!]
\centering
\caption{Covariates used in the marginal models. $i$ indexes the observations and $j$ indexes the macro regions.}
\label{tab:covariates}
{\small\renewcommand{\arraystretch}{0.85}%
\begin{tabularx}{\textwidth}{lX@{\hspace{1em}}lX}
\toprule
\multicolumn{2}{l}{\textbf{General covariates}} & \multicolumn{2}{l}{\textbf{Weather forecast covariates}} \\
\midrule
$dow_j$          & day of the week factor                                                      & $rain_{ij}$        & mean precipitation (m/day) \\
$dow_{ij}^{+}$   & $dow_j$ with additional factor levels for public holidays                    & $temp_{ij}$        & temperature (K) at cell with highest regional population density \\
$t_i$            & time since January 1st, 2014                                                & $temp_{ij}^{S}$    & 48 hr rolling mean of $temp_{ij}$ \\
$shol_{ij}$      & school holidays, three-level factor distinguishing Christmas from other holidays & $irr_{ij}$      & mean solar irradiance (W/m$^2$) times embedded solar capacity (MW) \\
$tod_j$          & time of day ($\in \{0,0.5,\dots,23.5\}$)                                    & $wsp_{ij}^{10}$    & mean wind speed at 10 meters (m/s) \\
$wcap_i$         & GB embedded wind generation capacity (MW)                                   & $wsp_{ij}^{100}$   & mean wind speed at 100 meters (m/s) \\
$doy_j$          & day of the year ($\in \{1,\dots,366\}$)                                     & & \\
$n2ex_i$         & N2EX day-ahead electricity price (\pounds/MWh)                              & & \\
$y_{ij}^{24}$    & net-demand at a 24 hr lag                                                   & & \\
\bottomrule
\end{tabularx}}
\end{table}

In particular, we consider multivariate probabilistic forecasting of net-demand in Great Britain using a Gaussian location–scale GAMLSS-type model for the marginals and a conditional vine copula model for the dependence structure. As \citet{gioia2025additive}, we use data from 2014 to 2018, where net-demand is observed at 30-minute resolution for 14 Grid Supply Point (GSP) groups. We aggregate the 14 GSP groups to five macro-regions ordered from north to south, as shown in Figure~\ref{fig:uk_map}, broadly following the regional groupings used by \citet{nationalgrid2018regional} to report demand and generation trends across Great Britain. Let \(j=1,\dots,d\), \(d=5\), index these macro-regions, and let \(\mathcal{G}_j\) denote the set of GSP groups assigned to region \(j\). The covariates used in the marginal models are reported in Table~\ref{tab:covariates} and the aggregation scheme is reported in Table~\ref{tab:aggregation} of Appendix~\ref{app:aggregation}. The weather variables are aggregated to the macro-regional level using means or sums, depending on the respective variable, while calendar variables are mapped to the respective macro-regions.

\begin{figure}[t!]
\centering
\input{figures/timeline_modes.tex}
\caption{Illustration of the three updating schemes evaluated in Table~\ref{tab:forecast_eval}.
  An orange dot marks a batch fit, at which the pair-copula families are selected and all parameters are re-estimated on the data available so far.
  A blue tick marks a sequential online update after a single observation.
  Shaded bars mark stretches in which the parameters are held fixed.
  Marginal models and copula follow the same scheme in each row.
  The test set spans twelve months; for clearer exposition the figure shows only three.}
\label{fig:timeline_modes}
\end{figure}
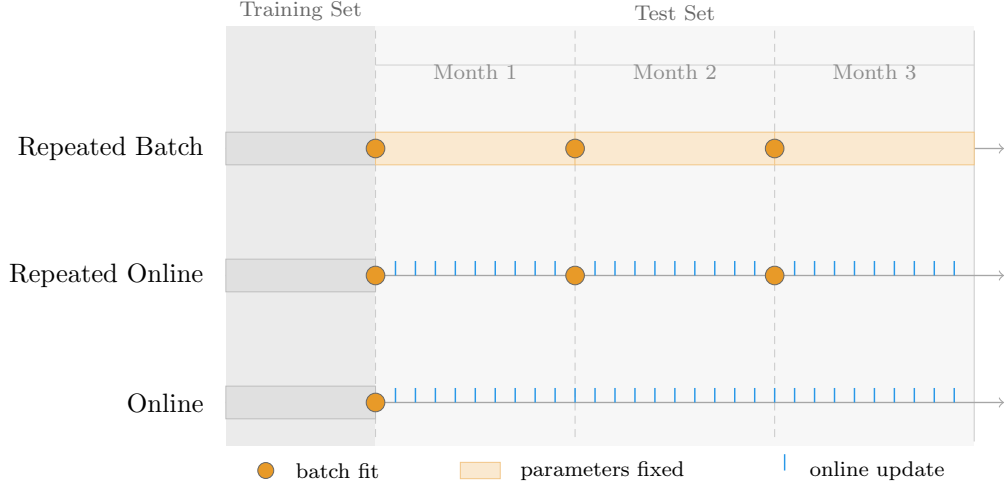

We emphasize again the key differences to \citet{gioia2025additive}. First, we allow for non-Gaussian dependence and second, we implement online estimation, i.e.,~we update the model parameters sequentially as new data becomes available. 
To isolate the contribution of the latter, we compare three updating schemes,
illustrated in Figure~\ref{fig:timeline_modes}. All three start from the same initial batch fit on the training data. In each of them the marginal models and the copula are estimated under the same scheme, i.e., if the copula is estimated purely online then also the marginal model is estimated purely online. Under the repeated batch scheme, the model is re-estimated on all available data at every monthly boundary, including the pair-copula family selection, and is held fixed within the month. This is exactly what \citet{gioia2025additive} do using their multivariate forecasting model based on the multivariate normal distribution. The repeated online scheme keeps these monthly batch fits but updates the parameters sequentially after every observation in between. The purely online setting is initialized at the start of the test period on the training data. After the initial fit the parameters are then updated sequentially throughout the test period. The copula families are never re-selected.

\subsection{Marginal Model}

We model each marginal distribution by a Gaussian location--scale GAMLSS-type model in which both, the location $\mu_{ij}$ and the scale $\sigma_{ij}$ of region $j$ at observation $i$, are driven by covariates and the two parameters are updated by cycling between them. Writing $\vartheta\in\{\mu,\sigma\}$ for a generic distribution parameter, each is tied to its own linear predictor $\eta_{ij}^{(\vartheta)}$ through a parameter-specific link $h_\vartheta$,
\[
h_\mu(\mu_{ij}) = \mu_{ij} = \eta_{ij}^{(\mu)},
\qquad
h_\sigma(\sigma_{ij}) = \log \sigma_{ij} = \eta_{ij}^{(\sigma)},
\]
i.e.,\ the identity link for the mean and the log link ensuring that the standard deviation is positive. Both predictors share the same covariate structure but carry their own coefficients. We therefore state it once for a linear predictor,
\begin{align*}
\eta_{ij} \;=\;
&\phantom{+}\underbrace{
   \begin{aligned}[t]
   &\phantom{+} p_{j1}(t_i)
   + p_{j2}(t_i^2)
   + p_{j3}(dow_{ij}^+)
   + p_{j4}(shol_{ij})
   + p_{j5}(y_{ij}^{24}) \\[2pt]
   &+ s_{j1}^{20}(doy_j)
   + s_{j2}^{35}(tod_j)
   + s_{j3}^{10}(n2ex_i)
   + s_{j4}^{35}(tod_j,\, dow_{ij}^+)
   + s_{j5}^{35}(tod_j,\, shol_{ij})
   \end{aligned}
   }_{\text{general covariates}} \\[10pt]
&\phantom{+}\underbrace{
   \begin{aligned}[t]
   &+ p_{j6}(wsp_{ij}^{10})
   + s_{j6}^{35}(temp_{ij})
   + s_{j7}^{35}(temp_{ij}^{S})
   + s_{j8}^{10}(rain_{ij}) \\[2pt]
   &+ s_{j9}^{5}(irr_{ij})
   + wcap_i \times s_{j10}^{20}(wsp_{ij}^{100}),
   \end{aligned}
   }_{\text{weather forecast covariates}}
\end{align*}
where $p_{jk}$ denotes a linear, parametric, effect of the respective covariate, while $s_{jk}^{K}$ denotes a smooth effect represented by a spline basis of dimension $K$ (given as the superscript), with $i$ indexing the observations and $j$ indexing the macro regions as before. Each smooth term is a basis expansion $s_{jk}^{K}(v) = \sum_{m=1}^{K} \beta_{jkm}^{(\vartheta)}\, B_{km}(v)$ in fixed spline basis functions $B_{k1},\dots,B_{kK}$, so that evaluating it at the observed covariates linearizes it into $K$ additional design columns. Collecting the parametric covariates together with all such basis evaluations for observation $i$ into a single design vector $\vx_{ij}$, the predictor is linear in the coefficients,
\[
\eta_{ij}^{(\vartheta)} = \vx_{ij}^{\top}\,\vbeta_{j}^{(\vartheta)},
\qquad \vartheta\in\{\mu,\sigma\}.
\]
The predictor is thus nonlinear in the covariates but linear in the coefficients $\vbeta_{j}^{(\vartheta)}$. Unlike a classical GAMLSS, we therefore do not penalize the roughness of the smooth effects, but control their flexibility through a ridge ($\ell_2$) penalty on all coefficients except the intercept. Hence, the marginal models fit within the same penalized GLM framework as the vine copulas in Section~\ref{sec:methodology} and can be estimated online using the approach in \citet{hirsch2024online}.

For each region, \(j=1,\dots,5\), we, thus, estimate a conditional marginal model and obtain the fitted conditional parameters \(\hat{\mu}_{j,i}\) and \(\hat{\sigma}_{j,i}\). Applying the probability integral transform, i.e.,\ evaluating the implied conditional distribution function at the realized net-demand $y_{j,i}$, then yields the pseudo-observations
\begin{equation}\label{eq:pseudo}
  u_{j,i} = F_{j,i}(y_{j,i} \mid \vx_i)
= \Phi\!\left(\frac{y_{j,i} - \hat{\mu}_{j,i}}{\hat{\sigma}_{j,i}}\right),
\qquad \vu_i = (u_{1,i},\dots,u_{5,i})^\top \in (0,1)^5,
\end{equation}
where $\Phi$ is the standard normal cumulative distribution function. These pseudo-observations are then used to estimate the conditional vine copula.

\subsection{Vine Copula}

We model the dependence structure using a D-vine copula that orders the variables according to their spatial connectedness (see Figure~\ref{fig:uk_map}). For each edge in every tree, we estimate a bivariate GLM for each of the five candidate families introduced in Section~\ref{sec:bivariate_copulas}, Gaussian, Clayton~I, Clayton~II, Gumbel~I, and Gumbel~II, and select the best-fitting one via the AIC. Regarding the covariate specifications, we distinguish between trees. In the first tree, where dependence is strongest, the copula parameter on each edge $e=(r,s)$ is allowed to vary with edge-specific weather covariates. In subsequent trees, the copula parameters are regressed on an intercept only. This is due to the fact that estimation gets distorted in higher trees as shown in our simulations in Section~\ref{sec:simulations}. The ridge penalty is applied only to the estimation in the first tree, leaving the intercept unpenalized. Throughout the application we set the forgetting factor to $\gamma=0$. As in Section~\ref{sec:bivariate_copulas}, each edge in the first tree carries a linear predictor $\eta_{e,i} = \vx_{e,i}^\top\vbeta_e$, which is mapped to Kendall's $\tau$ through the link $g(\eta) = \tanh(\eta/2)$ and then to the family-specific copula parameter through the bijection $f$,
\[
\tau_{e,i} = g(\eta_{e,i}),
\qquad
\theta_{e,i} = f(\tau_{e,i}) = f\!\left(g(\eta_{e,i})\right),
\]
with the edge-specific linear predictor given by
\begin{align*}
\eta_{e,i}
&= \underbrace{
   \beta_{e,0}
   + \beta_{e}^{\mathrm{temp}}\,\Delta \mathrm{temp}_{e,i}
   + \beta_{e}^{\mathrm{rain}}\,\Delta \mathrm{rain}_{e,i}
   + \beta_{e}^{\mathrm{irr}}\,\Delta \mathrm{irr}_{e,i}
   + \beta_{e}^{\mathrm{wsp10}}\,\Delta \mathrm{wsp}^{10}_{e,i}
   + \beta_{e}^{\mathrm{wsp100}}\,\Delta \mathrm{wsp}^{100}_{e,i}
   }_{\text{pair-specific}} \notag \\[10pt]
&\quad+\underbrace{
   \textstyle\sum_{k\in\{\text{Tue},\dots,\text{Sun}\}}\!\gamma_{e,k}\,\mathbf{1}[dow_i = k]
   + \delta_{e,1}\,\sin\!\left(\tfrac{2\pi\,tod_i}{24}\right)
   + \delta_{e,2}\,\cos\!\left(\tfrac{2\pi\,tod_i}{24}\right)
   }_{\text{non-pair-specific}}, 
\end{align*}
where $\Delta x_{e,i} = |x_{r,i} - x_{s,i}|$ is the absolute pairwise weather spread, $\mathbf{1}[dow_i=k]$ are day-of-week dummies with Monday as baseline (common to all edges and varying only with observation $i$), and $(\delta_{e,1},\delta_{e,2})$ encode time of day harmonically. Pairwise weather spreads between adjacent regions are 
natural covariates for time-varying dependence parameters. They indicate whether
neighbouring regions are exposed to a common or heterogeneous weather regime. Consider, for example, cold and
calm days: Heating demand rises while embedded wind generation falls across
several neighbouring regions, pushing regional net-demands jointly towards their upper
tail. Conversely, when weather spreads are pronounced, for example with strong
wind in one region but calm conditions in an adjacent region, regional net-demands may
partially decouple. This covariate-based structure is useful not only statistically, but also economically, because it links changes in the dependence structure to interpretable covariates.

Schematically, our vine copula forecasting approach proceeds as follows:
\begin{enumerate}
    \item For each edge \(e\) in the vine, compute the corresponding copula parameter using the estimated pseudo-observations from the marginal models in \eqref{eq:pseudo} and
    \[
    \theta_{e,n+1} = f\left(g(\eta_{e,n+1})\right),
    \]
    where \(\eta_{e,n+1}\) is evaluated at the covariates \(\vx_{n+1}\) (in the first tree based on regional weather spreads). This yields the conditional vine copula \(C_{n+1}(\cdot \mid \vx_{n+1})\).

    \item Draw \(M\) independent samples from the conditional vine copula,
    \[
    \vu_{n+1}^{(m)} \sim C_{n+1}(\cdot \mid \vx_{n+1}),
    \qquad m=1,\dots,M,\quad \vu_{n+1}^{(m)} \in (0,1)^d,
    \]
    by evaluating its inverse Rosenblatt transform at independent \(U(0,1)^d\) vectors, i.e.,\ through the inverse $h$-functions of the vine copula \citep{dissmann2013selecting}. 

    \item Transform $\vu_{n+1}$ into $\vy_{n+1}$ via the previously estimated marginal models
    \[
    y_{j,n+1}^{(m)} = F_{j,n+1}^{-1}\!\left(u_{j,n+1}^{(m)} \mid \vx_{n+1}\right),
    \qquad j=1,\dots,d.
    \]
\end{enumerate}

The resulting sample \(\{\vy_{n+1}^{(m)}\}_{m=1}^M\) approximates the multivariate predictive distribution of \(\vy_{n+1}\).

\subsection{Results}

\begin{table}[t!]
\centering
\begin{threeparttable}
\caption{Forecast evaluation under different updating schemes.}
\label{tab:forecast_eval}

\input{tables/tab_forecast_eval.tex}

\begin{tablenotes}[flushleft]
\footnotesize
\item Notes: ES denotes the energy score, VS$_{0.5}$ the variogram score with $p=0.5$, DSS the Dawid--Sebastiani score, and LS the log score. Lower values indicate better predictive performance.  Each row is a complete forecasting model, in which the marginal models and the copula are estimated under the same updating scheme. The black underlining highlights the best model for each score, and the green shading the best updating scheme within each copula.
\end{tablenotes}

\end{threeparttable}
\end{table}
Finally, we evaluate the forecasting performance of our marginal plus conditional vine copula (M+CVC) model against two unconditional dependence models, namely a marginal plus vine copula (M+VC) and a Gaussian copula (M+GC) model in Table~\ref{tab:forecast_eval}. We use the energy score \citep[ES;][]{gneiting2007strictly}, the variogram score \citep[VS;][]{scheuerer2015variogram}, the Dawid–Sebastiani score \citep[DSS;][]{dawid1999coherent}, and the log score \citep[LS;][]{gneiting2007strictly} to evaluate the performance of the models. 

Table~\ref{tab:forecast_eval} reports the average scores across the test period for all models and updating schemes. The black underlining highlights the best model for each score, and the green shading the best updating scheme within each class of models. A consistent ordering emerges. Across models, the M+CVC using repeated online updates attains the smallest score in every column (underlined) and within every model (M+GC, M+VC, M+CVC) the repeated online updating scheme attains the smallest score.

First, we further analyze the overall best forecasting model across copulas and updating schemes. We, thus, report the model confidence set (MCS) of \citet{hansen2011model} in Table~\ref{tab:mcs}. The procedure starts from all nine models and eliminates them sequentially, correcting for the multiplicity of the pairwise comparisons. What remains is the set of models that cannot be distinguished from the best one at a given confidence level.

The M+CVC under repeated online updating is the only model in the $90\%$ model confidence set, and it is so for every scoring rule. The closest competitors are the unconditional models under repeated online updating on the VS, with $p$-values of $0.068$ and $0.058$. They stay outside the set here, but enter it at the $95\%$ level, and also under a longer bootstrap block. The VS is the scoring rule on which the models are hardest to separate, consistent with the small differences already visible in Table~\ref{tab:forecast_eval}. On the remaining three scoring rules every competitor is eliminated with a $p$-value below $0.001$. We report pairwise Diebold--Mariano tests (DM tests) in Appendix~\ref{app:dm_decomposition} for completeness and confirm the results from the MCS.

\begin{table}[t!]
\centering
\begin{threeparttable}
\caption{Model confidence set $p$-values based on $T_{\max}$.}
\label{tab:mcs}

\input{tables/tab_mcs.tex}

\begin{tablenotes}[flushleft]
\footnotesize
\item Notes: Each cell reports the MCS $p$-value of \citet{hansen2011model}, computed separately for each scoring rule. We follow the settings of the R package \texttt{MCS}, that is a moving block bootstrap with $5000$ replications, the statistic $T_{\max}$, and an estimated block length of $42$ observations. $T_{\max}$ compares each model with the average of the surviving set. Models in bold belong to the $90\%$ model confidence set.
\end{tablenotes}

\end{threeparttable}
\end{table}
Two questions remain. First, does the M+CVC outperform its unconditional benchmarks when the updating scheme is held fixed? Second, does repeated online updating outperform its two competitors when the copula is held fixed? Appendix~\ref{app:dm_decomposition} answers both questions with pairwise DM tests that vary one factor (model or updating scheme) at a time. Table~\ref{tab:dm_within_scheme} holds the updating scheme fixed and compares the M+GC and the M+VC to the M+CVC. Most differences are significant at the $5\%$ level. The exception is the VS, where the evidence is only significant at the $10\%$ level for repeated online updating and not significant anymore under repeated batch estimation. This confirms the results from the MCS. Table~\ref{tab:dm_within_copula} holds the copula fixed and compares the updating schemes to repeated online updating. Here, repeated online updating proves to be statistically superior at the 1 \% level throughout. The answer to both questions is, thus, yes: the M+CVC outperforms its unconditional benchmarks within every updating scheme on ES, DSS and LS, with weaker evidence on the VS, and repeated online updating outperforms its competitors for every copula. 

Two remarks are in order. First, the differences between the M+GC and the M+VC are small. The vine selects Gaussian copulas at almost every monthly updating step, and only a single edge is estimated to be Gumbel (see Figure~\ref{fig:tau_all_edges} in Appendix~\ref{app:kendallstau}). The gain therefore rather comes from conditioning the dependence on covariates rather than from the vine construction itself. Second, fully online updating does not replace periodic re-estimation. Without the monthly refit the coefficients drift away from the periodically re-estimated fit over the test period. Forecasting performance is, thus, significantly worse than for the repeated online case. 
\begin{figure}[t!]
\centering
\includegraphics[width=\linewidth]{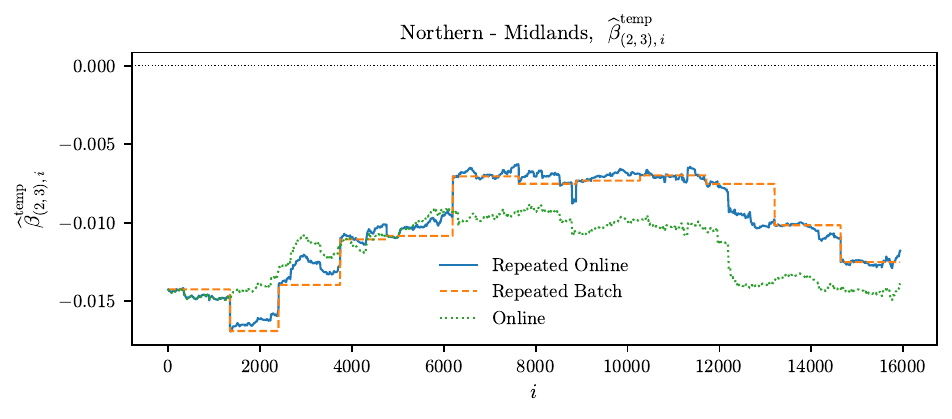}
\caption{Estimated coefficient on the Northern--Midlands temperature spread (first tree, CVC specification) over the test period separated by updating scheme. We write $\widehat{\beta}^{\mathrm{temp}}_{(2,3),i}$ for the temperature-spread slope of edge $e=(2,3)$ as estimated after observation $i$.}
\label{fig:beta_zoom_online}
\end{figure}
Figure~\ref{fig:beta_zoom_online} makes that drift visible for the Northern--Midlands temperature-spread coefficient, $\hat{\beta}^{\mathrm{temp}}_{(2,3),i}$, in the first tree of the CVC. The repeated batch estimates (orange) remain constant between monthly boundaries, while the repeated online estimates (blue) adjust within each month and reset at the next boundary. The online estimates (green), meanwhile, follow the same broad pattern but deviate more clearly over some periods, especially from about $i=6000$ onwards. In our setting, online updating is therefore best understood as a complement to periodic re-estimation rather than a substitute for it. Graphs for the other covariates and for all other edges look qualitatively similar and are available upon request.

Figure~\ref{fig:beta_zoom_online} can also be interpreted economically. The results show that the temperature-spread is an important driver of dependence. A large spread in temperature between two adjacent regions is associated with weaker dependence between their net-demands. This is consistent with the physical intuition that heterogeneous weather conditions can decouple regional net-demands, while common weather regimes align them.

We, additionally, report computation times for the different updating schemes in Table~\ref{tab:computation_times} of Appendix~\ref{app:computationtimesapp}. An average online update takes under a second for the Gaussian and unconditional vine copulas and around two to three seconds for the conditional vine copula, whereas a full batch refit ranges from under ten seconds for the Gaussian copula to about 190 seconds for the conditional vine copula. The comparison is not totally fair in the sense that the batch procedure is re-estimating the model using copula selection, but it emphasizes again that online estimation allows for more frequent updates at a fraction of the computational cost.

\section{Conclusion}\label{sec:conclusion}

We propose an online, regularized GLM framework for conditional vine copulas, in which each pair-copula parameter is driven by covariates through a Kendall's $\tau$ link. The model is estimated by a ridge-penalized online algorithm that updates parameters sequentially as new data arrive. Simulations confirm that the estimator recovers the true coefficients and that online updating matches repeated batch refitting at a fraction of the computational cost. Applying the model to joint regional electricity net-demand forecasting in Great Britain, we find that the conditional vine copula model using repeated online updating is the only model in the $90\%$ model confidence set, and it is so for every scoring rule. Decomposing the comparison shows that the vine construction alone does improve on a Gaussian copula only mildly but conditioning it on weather-spread covariates yields significant gains across several scoring rules. In our setting, online updating is useful as a complement to, rather than a substitute for, periodic re-estimation.

\section*{Data and Code Availability}

The replication package contains the code for both the simulation study and the empirical application, together with the instructions needed to reproduce every table and figure of this paper, and is available at \url{https://github.com/chrischu12/online-vine-copulas}.
The estimation algorithm itself is implemented in the open source \texttt{Python} package \texttt{ondil}. The vine copula machinery builds on the \texttt{VineCopulas} package of \citet{claassen2024vinecopulas}. The regional net-demand data are not redistributed with the package; they are taken from the replication material of \citet{gioia2025additive}, available at \url{https://doi.org/10.5281/zenodo.13330081}.

\section*{Generative AI Statement}

During the preparation of this work, the authors used generative AI tools, including ChatGPT and Claude Code, to support general manuscript review and language editing, and assist with coding tasks. The latter included, in particular, code review, documentation, and the creation of figures and visualizations. All AI-assisted output was reviewed, verified, and edited by the authors as necessary. The authors take full responsibility for the content of the publication.

\newpage

\singlespacing

\bibliographystyle{apalike}
\bibliography{references}

\newpage

\section*{Appendix}
\begin{appendix}

  \section{Methodological Details}\label{app:derivations}

This appendix provides the derivations underlying the estimation procedure. We recast the Newton--Raphson update for the penalized copula log-likelihood as an iteratively reweighted ridge regression (\ref{app:derivations_newton_raphson}), derive the Gaussian copula score and Hessian in closed form (\ref{app:normal_copula}), and provide additional details on the OCD algorithm (\ref{app:ocd}). 

  \subsection{Newton--Raphson Update as Iteratively Reweighted Ridge Regression}\label{app:derivations_newton_raphson}
For $\vu\in[0,1]^2$ and copula parameter $\theta\in\Theta$, define
\[
\ell(\vu,\vx;\vbeta) = \log c\!\left(\vu;\theta(\vx,\vbeta)\right),
\qquad \theta(\vx,\vbeta) = f\!\left(g(\vx^\top\vbeta)\right).
\]
Given a sample $\{(\vu_i,\vx_i)\}_{i=1}^n$, the penalized log-likelihood is
\[
\ell_c(\vbeta) = \ell(\vbeta) - \frac{1}{2}\lambda\,\vbeta^\top\vbeta,
\qquad
\ell(\vbeta)=\sum_{i=1}^n \ell(\vu_i,\vx_i;\vbeta)=: \sum_{i=1}^n \ell_i,
\]
and the penalized maximum likelihood estimator reads
\[
\widehat{\vbeta}_\lambda = \arg\max_{\vbeta\in\SR^p} \ell_c(\vbeta).
\]

We show in the following that the Newton--Raphson update for $\widehat{\vbeta}_\lambda$ can be reformulated as a weighted ridge regression problem. We, thus, derive the score, $s_i(\vbeta)$, and the weights,  $-w_i(\vbeta)$, of $\ell_c$ with respect to $\veta$ and plug these results into the Newton--Raphson update formula. 

For each $i$, $\ell_i$ depends on $\vbeta$ only through the scalar $\eta_i = \vx_i^\top\vbeta$. 
Applying the chain rule through $\eta_i \mapsto \tau_i \mapsto \theta_i$,
\begin{equation}
\frac{\partial \ell_i}{\partial \eta_i}
= \frac{\partial \ell_i}{\partial \theta_i}\, f'(\tau_i)\, g'(\eta_i)
\;=:\; s_i(\vbeta)\,
\label{eq:score_eta}
\end{equation}
and since $\partial \eta_i / \partial \vbeta = \vx_i$,
\[
\frac{\partial \ell_i}{\partial \vbeta} = s_i(\vbeta)\,\vx_i.
\]
Differentiating $s_i$ once more with respect to $\eta_i$ gives the full second derivative
\begin{equation}
\frac{\partial^2 \ell_i}{\partial \eta_i^2}
= \frac{\partial^2 \ell_i}{\partial \theta_i^2}\,[f'(\tau_i)]^2\,[g'(\eta_i)]^2
+ \frac{\partial \ell_i}{\partial \theta_i}\!\left\{f''(\tau_i)\,[g'(\eta_i)]^2 + f'(\tau_i)\,g''(\eta_i)\right\}
\;=:\; -w_i(\vbeta),
\label{eq:weight}
\end{equation}
so that
\[
\frac{\partial^2 \ell_i}{\partial \vbeta\,\partial \vbeta^\top}
= -w_i(\vbeta)\,\vx_i\vx_i^\top.
\]

Now, collect $\vs(\vbeta) = (s_1(\vbeta),\ldots,s_n(\vbeta))^\top$ and $\mW(\vbeta) = \operatorname{diag}(w_1(\vbeta),\ldots,w_n(\vbeta))$. Then
\[
\nabla \ell_c(\vbeta) = \mX^\top \vs(\vbeta) - \lambda\vbeta,
\qquad
\nabla^2 \ell_c(\vbeta) = -\mX^\top \mW(\vbeta)\,\mX - \lambda\mI.
\]

The Newton--Raphson update $\widehat{\vbeta}^{[t+1]} = \widehat{\vbeta}^{[t]} - [\nabla^2 \ell_c]^{-1}\nabla \ell_c$, evaluated at $\widehat{\vbeta}^{[t]}$, reads
\[
\widehat{\vbeta}^{[t+1]}
= \widehat{\vbeta}^{[t]} + \left[\mX^\top \mW^{[t]} \mX + \lambda \mI\right]^{-1}\!\left[\mX^\top \vs^{[t]} - \lambda\widehat{\vbeta}^{[t]}\right],
\]
or equivalently
\[
\left[\mX^\top \mW^{[t]} \mX + \lambda \mI\right]\!\bigl(\widehat{\vbeta}^{[t+1]} - \widehat{\vbeta}^{[t]}\bigr)
= \mX^\top \vs^{[t]} - \lambda\widehat{\vbeta}^{[t]}.
\]
Adding $[\mX^\top \mW^{[t]} \mX + \lambda \mI]\widehat{\vbeta}^{[t]}$ to both sides collapses the left-hand side to $[\mX^\top \mW^{[t]} \mX + \lambda \mI]\widehat{\vbeta}^{[t+1]}$, while on the right we expand $[\mX^\top \mW^{[t]} \mX + \lambda \mI]\widehat{\vbeta}^{[t]} = \mX^\top \mW^{[t]} \mX\widehat{\vbeta}^{[t]} + \lambda\widehat{\vbeta}^{[t]}$, such that
\[
\left[\mX^\top \mW^{[t]} \mX + \lambda \mI\right]\widehat{\vbeta}^{[t+1]}
= \mX^\top \vs^{[t]} + \mX^\top \mW^{[t]} \mX\widehat{\vbeta}^{[t]},
\]
factoring out $\mX^\top$,
\[
\left[\mX^\top \mW^{[t]} \mX + \lambda \mI\right]\widehat{\vbeta}^{[t+1]}
= \mX^\top\!\left(\mW^{[t]}\mX\widehat{\vbeta}^{[t]} + \vs^{[t]}\right)
= \mX^\top \mW^{[t]}\, \vz^{[t]},
\]
with working vector
\[
\vz^{[t]} := \veta^{[t]} + \bigl(\mW^{[t]}\bigr)^{-1}\vs^{[t]},
\qquad \veta^{[t]} = \mX\widehat{\vbeta}^{[t]},
\] such that
\[
\widehat{\vbeta}^{[t+1]} = \left[\mX^\top \mW^{[t]} \mX + \lambda \mI\right]^{-1} \mX^\top \mW^{[t]}\, \vz^{[t]}.
\]
Equivalently, $\widehat{\vbeta}^{[t+1]}$ solves the weighted ridge problem
\[
\widehat{\vbeta}^{[t+1]} = \arg\min_{\vbeta \in \mathbb{R}^p}\left\{\bigl\|\vz^{[t]} - \mX\vbeta\bigr\|_{\mW^{[t]}}^2 + \lambda\|\vbeta\|_2^2\right\},
\qquad \|\vu\|_{\mW}^2 := \vu^\top \mW \vu.
\]

The derivations so far are generic in the link $g$, the bijection $f$, and the copula density $c$. For all three families in our application the link $g$ from the linear predictor to Kendall's $\tau$ is the Fisher-$z$ link
\[
\tau = g(\eta) = \tanh\!\Big(\tfrac{\eta}{2}\Big),
\qquad
g'(\eta) = \tfrac12\operatorname{sech}^2\!\Big(\tfrac{\eta}{2}\Big),
\qquad
g''(\eta) = -\tfrac12\operatorname{sech}^2\!\Big(\tfrac{\eta}{2}\Big)\tanh\!\Big(\tfrac{\eta}{2}\Big),
\]
so the families differ only in the bijection $f:(-1,1)\to\Theta$ from $\tau$ to the native parameter $\theta$ and in the log-density $\ell=\log c$. Since $\ell=\log c$, the $\theta$-derivatives entering the score \eqref{eq:score_eta} and weight \eqref{eq:weight} are
\begin{equation}
\frac{\partial\ell}{\partial\theta}=\frac{1}{c}\frac{\partial c}{\partial\theta},
\qquad
\frac{\partial^2\ell}{\partial\theta^2}=\frac{1}{c}\frac{\partial^2 c}{\partial\theta^2}-\left(\frac{\partial\ell}{\partial\theta}\right)^{2}.
\label{eq:logdens_identity}
\end{equation}
We work out the Normal copula in detail below.

\subsection{Normal Copula}\label{app:normal_copula}

The dependence parameter $\theta\in(-1,1)$ is the Pearson correlation, and the bijection from Kendall's $\tau$ is (Table~\ref{tab:copula_families})
\[
\theta = f(\tau) = \sin\!\Big(\tfrac{\pi}{2}\tau\Big),
\qquad
f'(\tau) = \tfrac{\pi}{2}\cos\!\Big(\tfrac{\pi}{2}\tau\Big),
\qquad
f''(\tau) = -\Big(\tfrac{\pi}{2}\Big)^{2}\sin\!\Big(\tfrac{\pi}{2}\tau\Big).
\]
With $\Phi$ the standard normal CDF, so that $\Phi^{-1}(u_1)$ and $\Phi^{-1}(u_2)$ are the normal quantiles of the pseudo-observations, the log-density is
\begin{equation*}
\ell(u_1,u_2;\theta) = -\tfrac12\log(1-\theta^2) + \frac{2\theta\,\Phi^{-1}(u_1)\Phi^{-1}(u_2) - \theta^2\bigl(\Phi^{-1}(u_1)^2+\Phi^{-1}(u_2)^2\bigr)}{2(1-\theta^2)},
\end{equation*}
with score
\begin{equation*}
\frac{\partial \ell}{\partial \theta}
= \frac{\theta(1-\theta^2) - \theta\bigl(\Phi^{-1}(u_1)^2+\Phi^{-1}(u_2)^2\bigr) + (1+\theta^2)\Phi^{-1}(u_1)\Phi^{-1}(u_2)}{(1-\theta^2)^2}
\end{equation*}
and curvature
\begin{equation*}
\frac{\partial^2 \ell}{\partial \theta^2}
= \frac{(1-\theta^2)(1+\theta^2) + 2\theta(\theta^2+3)\Phi^{-1}(u_1)\Phi^{-1}(u_2) - (1+3\theta^2)\bigl(\Phi^{-1}(u_1)^2+\Phi^{-1}(u_2)^2\bigr)}{(1-\theta^2)^3}.
\end{equation*}

The Gumbel and Clayton copulas are handled in exactly the same way. Only the bijection $f$ (Table~\ref{tab:copula_families}) and the log-density $\ell=\log c$ differ, while the Fisher-$z$ link $g$, the chain rule \eqref{eq:score_eta}--\eqref{eq:weight}, and the identity \eqref{eq:logdens_identity} carry over unchanged. Their scores and curvatures follow from the same differentiation, with negative $\tau$ covered by the $90^\circ$ and $270^\circ$ rotations of Equations~\eqref{eq:family_rotation_i}--\eqref{eq:family_rotation_ii}. The explicit expressions are documented in the \texttt{ondil} package.

\subsection{Online Coordinate Descent}\label{app:ocd}

This appendix gives the coordinate descent step in line~\ref{line:ocd} of Algorithm~\ref{alg:online_update}. For a given $\lambda\in\Lambda$, the ridge-penalized Newton--Raphson update solves the linear system in \eqref{eq:ridge_closed_form},
\begin{equation*}
    \widehat{\vbeta}_{N+1,\lambda}^{[t+1]} = \Bigl(\mG_{N+1}^{[t]} + \lambda\,\mI\Bigr)^{-1}\vh_{N+1}^{[t]},
\end{equation*}
with $\mG_{N+1}^{[t]}$ and $\vh_{N+1}^{[t]}$ from \eqref{eq:gramian_update}. Instead of inverting this matrix for every $\lambda$, coordinate descent solves the system one coefficient at a time \citep{friedman2007pathwise,angelosante2010online,messner2019online}. \\

\begin{algorithm}[H]
  \caption{Online coordinate descent for the ridge path, see \citet{angelosante2010online} and \citet{messner2019online}}\label{alg:ocd_ridge}
  \DontPrintSemicolon
  \KwIn{Statistics $\mG_{N+1}^{[t]}$ and $\vh_{N+1}^{[t]}$ from \eqref{eq:gramian_update}, penalty grid $\Lambda$, current estimates $\widehat{\vbeta}_{N+1,\lambda}^{[t]}$ for $\lambda\in\Lambda$ (equal to $\widehat{\vbeta}_{N,\lambda}$ at $t=0$)}
  \For{$\lambda \in \Lambda$}{
    $\vbeta \gets \widehat{\vbeta}_{N+1,\lambda}^{[t]}$\;
    \Repeat{$\vbeta$ has converged}{
      \For{$j = 1, \ldots, p$}{
        Update $\beta_j$ by \eqref{eq:ridge_update_online}\;
      }
    }
    $\widehat{\vbeta}_{N+1,\lambda}^{[t+1]} \gets \vbeta$\;
  }
  \KwOut{$\widehat{\vbeta}_{N+1,\lambda}^{[t+1]}$ for all $\lambda \in \Lambda$}
\end{algorithm}

\newpage
\noindent Write $\mG=\mG_{N+1}^{[t]}$ and $\vh=\vh_{N+1}^{[t]}$. Solving row $j$ of the system for $\beta_j$, with all other coefficients at their current values, gives
\begin{equation}
    \beta_j \leftarrow \frac{h_j - \sum_{i\neq j} G_{ji}\,\beta_i}{G_{jj} + \lambda},
    \qquad j = 1,\dots,p,
    \label{eq:ridge_update_online}
\end{equation}
where $p$ is the number of coefficients in $\vbeta$. This is the ridge case of \citet[eq.~(16)]{friedman2007pathwise}. Algorithm~\ref{alg:ocd_ridge} summarizes the step; $\lambda^\ast$ is then selected by the AIC in \eqref{eq:ic}.

\section{Additional Simulation Results}\label{app:additional_simulation_results}

This appendix reports additional simulation results. We extend Section~\ref{sec:simulation_batch} through reporting the average vine estimation time per update under unregularized (OLS) estimation (\ref{app:computationtimesim}). Additionally, we complement Section~\ref{sec:online_vs_batch} and show how the forgetting factor $\gamma$ works (\ref{app:structuralbreak}).

\subsection{Computation Times}\label{app:computationtimesim}
We show the computation times for the unregularized (OLS) case. As in the ridge case, online updating is faster than refitting the vine copula from scratch. Updates take around 0.5 seconds per step on average, whereas the batch fit ranges from about 0.8 to 1.7 seconds depending on the forgetting factor. Both schemes are also considerably faster than their ridge counterparts, since the unregularized fit avoids evaluating the $\lambda$ grid of the regularization path.
\begin{table}[H]
    \centering
\caption{Average vine estimation time per update without regularization.}
\label{tab:time_ols}
\input{tables/table_vine_time_ols_avg.tex}
	\captionsetup[table]{font=small} 
	\caption*{\textit{Notes:} We compute the average estimation time per update across all observations in the test set in seconds. The batch vine is estimated from scratch after each new observation, while the online vine is updated sequentially. The forgetting factor $\gamma$ controls the window size applied to the vine estimation.}
\end{table}

\subsection{Structural Break in the DGP}\label{app:structuralbreak}

\begin{figure}[H]
\centering
\includegraphics[width=\linewidth]{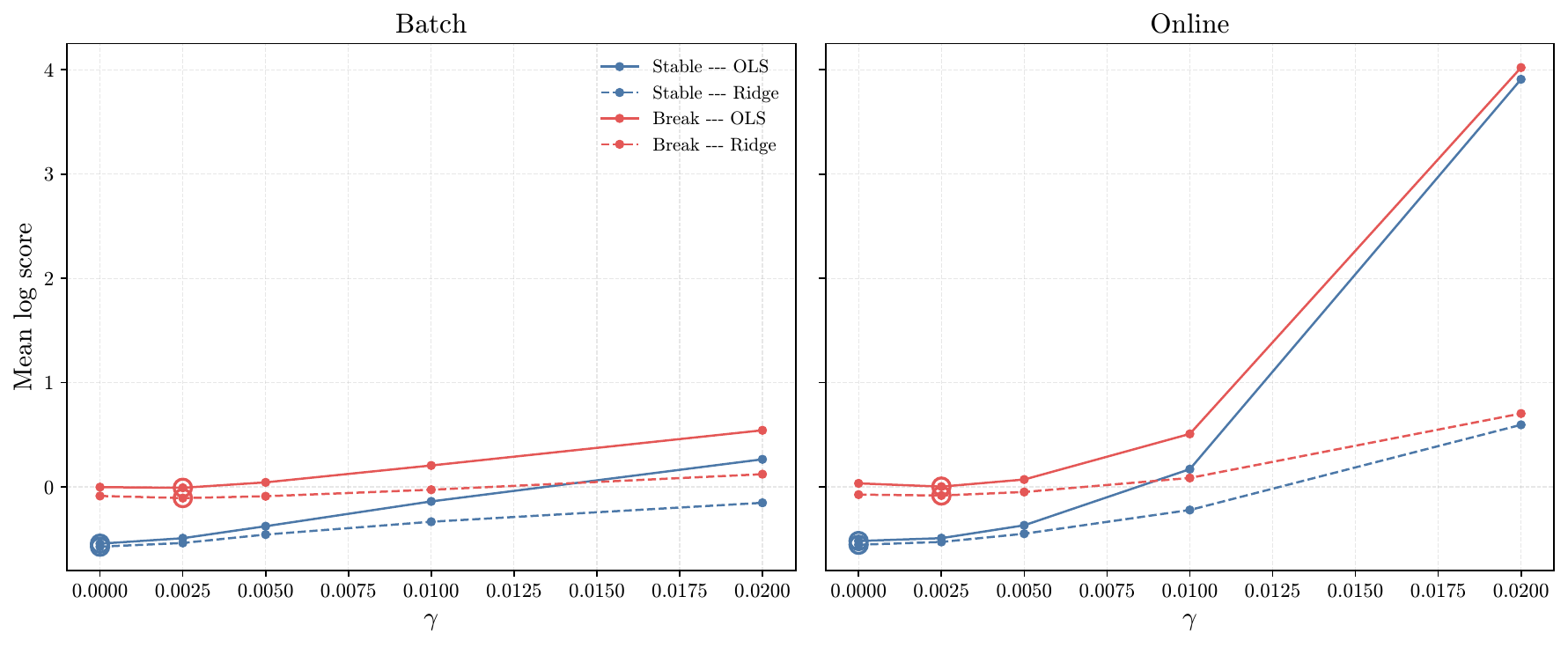}
\caption{Mean predictive log score as a function of the forgetting factor $\gamma$, for the stable and break DGPs. The left panel uses batch estimation, the right panel online estimation; solid lines are the unregularized (OLS) fit and dashed lines the ridge fit. Circles mark the score-minimizing $\gamma$. Lower is better.}
\label{fig:break_vs_stable_gamma}
\end{figure}

We illustrate the role of the forgetting factor $\gamma$ by comparing two scenarios. The first is the stable scenario of Section~\ref{sec:online_vs_batch}, in which the copula coefficients are constant over the sample. The second is a structural break scenario, identical to the stable one except that the first coefficient in the first tree is shifted from $0.25$ to $-0.25$ at observation $750$. This shift occurs shortly after the start of the online test period, so that the true dependence structure changes abruptly. Figure~\ref{fig:break_vs_stable_gamma} reports the mean predictive LS as a function of $\gamma$, separately for batch and online estimation and for the unpenalized and ridge estimation procedures.

Under the stable scenario the score is minimized at no forgetting ($\gamma = 0$). With a constant dependence structure it is simply the effective sample size that matters and, thus, using all past observations is optimal. Under the break process the optimum shifts to a mild degree of forgetting ($\gamma = 0.0025$). Lightly discounting older observations lets the algorithm adapt to the post-break dependence and attains a lower score than estimation based on all previous observations. Forgetting is thus beneficial precisely when the DGP is not stable.

\section{Additional Application Results}\label{app:additional_application_details}

This appendix collects additional results for the electricity net-demand application. We report the scheme used to aggregate the GSP-level series to the five macro regions (\ref{app:aggregation}), detailed computation times for the three updating schemes (\ref{app:computationtimesapp}), the model confidence set under the range statistic (\ref{app:mcs_range}), a decomposition of the forecast comparison of Table~\ref{tab:dm_vs_best} (\ref{app:dm_decomposition}), and the estimated Kendall's $\tau$ paths for every vine copula edge (\ref{app:kendallstau}).

\subsection{Aggregation to Macro Regions}\label{app:aggregation}

\begin{table}[!htpb]
\centering
\caption{Aggregation from GSP level to macro regions.}
\label{tab:aggregation}
\resizebox{\textwidth}{!}{%
\begin{tabular}{lll}
\hline
\textbf{Variable} & \textbf{Aggregation} & \textbf{Formula} \\
\hline

Net-demand $y_{ij}$
& Sum
& $y_{ij} = \sum_{g \in \mathcal{G}_j} y_{ig}$ \\

Wind speed $wsp_{ij}^{h}$
& Mean
& $wsp_{ij}^{h} = \frac{1}{|\mathcal{G}_j|} \sum_{g \in \mathcal{G}_j} wsp_{ig}^{h}$, \ $h \in \{10,100\}$ \\

Wind capacity $wcap_{ij}$
& Sum
& $wcap_{ij} = \sum_{g \in \mathcal{G}_j} wcap_{ig}$ \\

Solar proxy $irr_{ij}$
& Sum
& $irr_{ij} = \sum_{g \in \mathcal{G}_j} irr_{ig}$ \\

Temperature / precipitation
& Mean
& $x_{ij} = \frac{1}{|\mathcal{G}_j|} \sum_{g \in \mathcal{G}_j} x_{ig}$, \ $x \in \{temp, temp^{S}, rain\}$ \\

Calendar \& system vars 
& Representative GSP 
& $t_i,\, doy_j,\, tod_j,\, dow_j,\, dow_{ij}^{+},\, shol_{ij},\, n2ex_i$ \\

\hline
\end{tabular}%
}
\end{table}

Table~\ref{tab:aggregation} summarizes how the 14 GSP-level series are mapped to the five macro regions. Net-demand, wind capacity, and the solar proxy are summed over the GSP groups in each region. The weather variables, wind speed, temperature, and precipitation are averaged, and calendar and system variables, which are common within a region, are taken from a representative GSP.

\subsection{Computation Times}\label{app:computationtimesapp}

Table~\ref{tab:computation_times} separates the cost of the batch fits, which include copula selection, from that of the sequential online updates. A single online update is fast (under a second for the Gaussian and unconditional vine copulas and around two to three seconds for the conditional vine copula). Especially, it is much faster than a full batch refit, which takes from roughly ten seconds for the Gaussian copula to about 190 seconds for the conditional vine copula. Because the online schemes apply an update at each of the $15{,}936$ test steps, however, their cumulative cost over the whole test period exceeds that of the twelve monthly batch refits; the advantage of online updating is therefore within-month adaptation rather than a lower total computation time in our application.

\begin{table}[!htpb]
\centering
\begin{threeparttable}
\caption{Computation times of the copula-based models under different updating schemes.}
\label{tab:computation_times}

\input{tables/tab_computation_times.tex}

\begin{tablenotes}[flushleft]
\footnotesize
\item Notes: The \emph{Fit} block reports the number of batch fits ($N$), their average and total computation time in seconds; the \emph{Update} block the analogous quantities for the sequential online updates over the test period. The final column is the overall (fit + update) time. A dash indicates that the corresponding quantity is not applicable to the updating scheme.
\end{tablenotes}

\end{threeparttable}
\end{table}

\subsection{The Model Confidence Set under the Range Statistic}\label{app:mcs_range}

Table~\ref{tab:mcs} uses the statistic $T_{\max}$. The range statistic, $T_R$, on the other hand, considers all pairwise loss differences and takes the largest standardized one. With nine models, it therefore compares 36 pairs and can be driven by a single unusually large difference. Table~\ref{tab:mcs_tr} shows the model confidence set under the range statistic $T_R$. 

\begin{table}[H]
\centering
\begin{threeparttable}
\caption{Model confidence set $p$-values based on $T_R$.}
\label{tab:mcs_tr}

\input{tables/tab_mcs_tr.tex}

\begin{tablenotes}[flushleft]
\footnotesize
\item Notes: as in Table~\ref{tab:mcs}, but with the range statistic $T_R$ in place of $T_{\max}$. Models in bold belong to the $90\%$ model confidence set.
\end{tablenotes}

\end{threeparttable}
\end{table}

The MCS does not depend on the choice of  $T_{\max}$ or $T_R$. Under both statistics the M+CVC with repeated online updating is the only model in the $90\%$ set, on each of the four scoring rules. The two statistics differ only in the $p$-values of the unconditional models under repeated online updating on the VS, which stay below the threshold either way.

\subsection{Decomposing the Forecast Results}\label{app:dm_decomposition}

Table~\ref{tab:dm_vs_best} tests every model against the best one, the M+CVC under repeated online updating. This, however, does not tell us whether the differences are due to the updating scheme or the copula specification. The two tables that follow vary one factor at a time.

 Table~\ref{tab:dm_within_scheme} holds the updating scheme fixed and compares the M+GC and the M+VC to the M+CVC. Table~\ref{tab:dm_within_copula} holds the copula fixed and compares the updating schemes to repeated online updating.
\begin{table}[H]
\centering
\begin{threeparttable}
\caption{Two-sided \citet{diebold1995predictive} tests: General forecasting performance.}
\label{tab:dm_vs_best}

{\small\input{tables/tab_dm_vs_best.tex}}

\begin{tablenotes}[flushleft]
\footnotesize
\item Notes: Each cell reports the statistic of a two-sided DM test. The variance of the loss differential is estimated by the \citet{newey1987simple} estimator with 48 lags (one day). The loss differential is the score of the respective model minus the score of the M+CVC under repeated online updating, so a positive value means that the competing model forecasts worse. Stars denote significance at the $1$\%, $5$\% and $10$\% level.
\end{tablenotes}

\end{threeparttable}
\end{table}

\begin{table}[H]
\centering
\begin{threeparttable}
\caption{Two-sided \citet{diebold1995predictive} tests: Comparing copula specifications within the same updating scheme.}
\label{tab:dm_within_scheme}

{\small\input{tables/tab_dm_within_scheme.tex}}

\begin{tablenotes}[flushleft]
\footnotesize
\item Notes: Each cell reports the statistic of a two-sided DM test. The variance of the loss differential is estimated by the \citet{newey1987simple} estimator with 48 lags (one day). The loss differential is the score of the respective copula minus the score of the M+CVC under the same updating scheme, so a positive value means that the competing copula forecasts worse. Stars denote significance at the $1$\%, $5$\% and $10$\% level.
\end{tablenotes}

\end{threeparttable}
\end{table}

\begin{table}[H]
\centering
\begin{threeparttable}
\caption{Two-sided \citet{diebold1995predictive} tests: Comparing updating schemes across the same copula specification.}
\label{tab:dm_within_copula}

{\small\input{tables/tab_dm_within_copula.tex}}

\begin{tablenotes}[flushleft]
\footnotesize
\item Notes: Each cell reports the statistic of a two-sided DM test. The variance of the loss differential is estimated by the \citet{newey1987simple} estimator with 48 lags (one day). The loss differential is the score of the respective updating scheme minus the score of repeated online updating for the same copula, so a positive value means that the competing scheme forecasts worse. Stars denote significance at the $1$\%, $5$\% and $10$\% level.
\end{tablenotes}

\end{threeparttable}
\end{table}

\subsection{Copula Selection and Kendall's $\tau$ over Time}\label{app:kendallstau}

\begin{figure}[H]
\centering
\includegraphics[width=\linewidth]{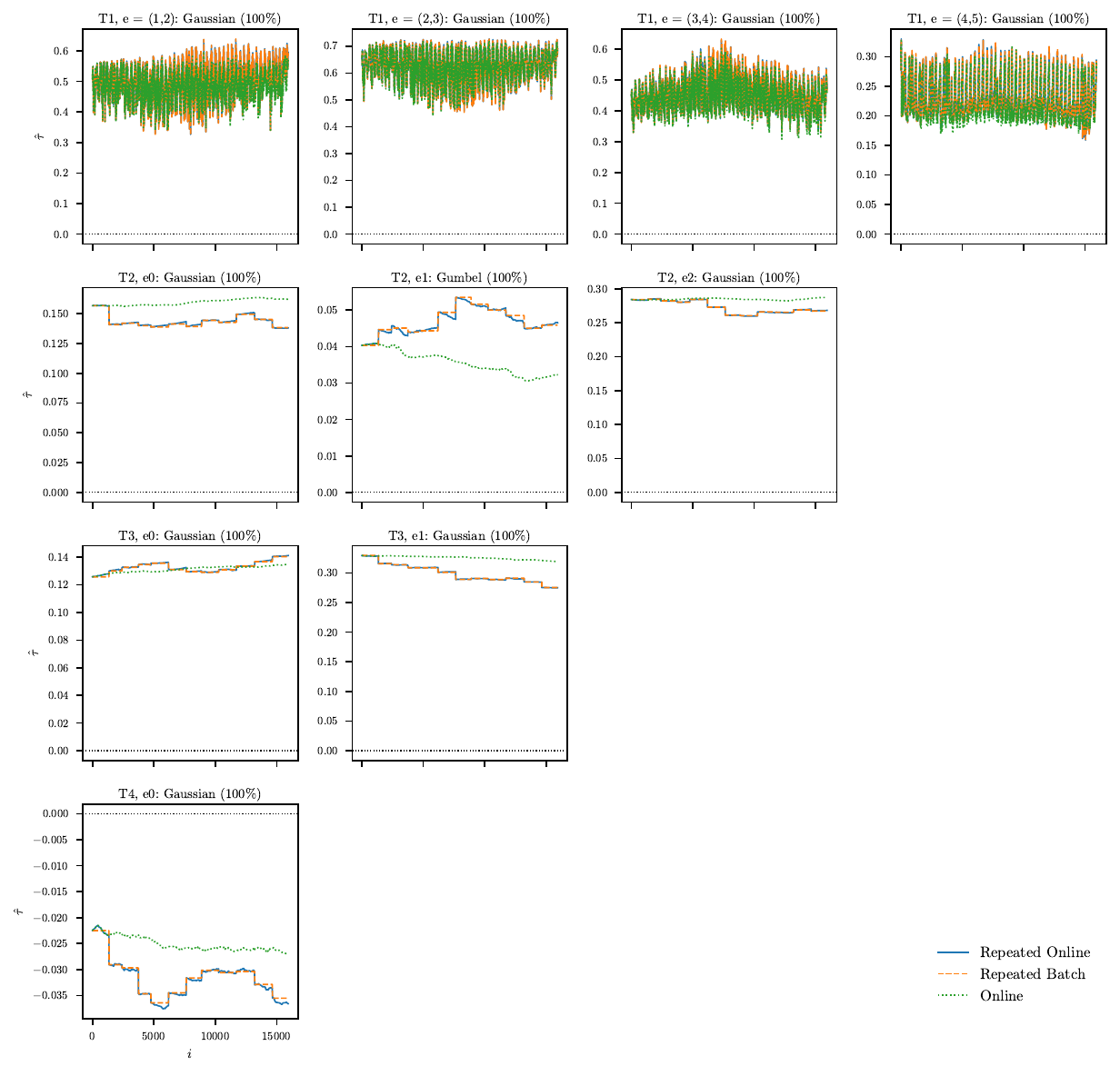}
\caption{Estimated Kendall's $\hat\tau$ over the test period for every pair-copula of the fitted CVC vine, under the three updating schemes. Each panel is one edge; its selected family and the share of test steps for which that family is chosen appear in the panel title. In the first tree, panels are labelled by the two macro regions they connect, in the higher trees by tree and edge number. }
\label{fig:tau_all_edges}
\end{figure}

Figure~\ref{fig:tau_all_edges} traces the estimated Kendall's $\tau$ for all $d(d-1)/2 = 10$ pair-copulas for the CVC specification. The dependence is strongest and most time-varying in the first tree, where the weather-spread covariates drive it, and weakens in the higher trees that are estimated with an intercept only. The selected family is almost always Gaussian, in line with the near-Gaussian dependence discussed in Section~\ref{sec:applications}.

\end{appendix}

\end{document}

%% file: tables/table_vine_time_ridge_avg.tex
\begin{tabular}{lrrrrr}
\toprule
 & $\gamma=0$ & $\gamma=1/400$ & $\gamma=1/200$ & $\gamma=1/100$ & $\gamma=1/50$ \\
\midrule
Batch & 21.452 & 19.542 & 16.721 & 13.153 & 10.405 \\
Online & 2.119 & 2.031 & 2.022 & 1.966 & 2.439 \\
\bottomrule
\end{tabular}

%% file: figures/uk_map.tex
\begin{tikzpicture}[
  every node/.style={font=\scriptsize},
  region/.style={draw=black!55, line width=0.25pt},
  letterlbl/.style={font=\tiny\bfseries, inner sep=0.3pt},
  macrolbl/.style={circle, fill=white, fill opacity=0.85, text opacity=1, draw=black!40, line width=0.2pt, inner sep=0.5pt, minimum size=4mm, font=\bfseries},
]
  \definecolor{r1}{HTML}{D85868}
  \definecolor{r2}{HTML}{68C858}
  \definecolor{r3}{HTML}{28E8E8}
  \definecolor{r4}{HTML}{E89A28}
  \definecolor{r5}{HTML}{2898E8}
  \definecolor{r1d}{HTML}{8B2230}
  \definecolor{r2d}{HTML}{1F7A1F}
  \definecolor{r3d}{HTML}{0C7A82}
  \definecolor{r4d}{HTML}{8C5500}
  \definecolor{r5d}{HTML}{0F4A8A}

  \filldraw[region, fill=r1]
    (0.997,4.081) -- (0.931,4.081) -- (0.887,4.135) -- (0.875,4.227) -- (0.927,4.295) -- (0.982,4.259) -- (0.981,4.190) -- (1.018,4.139) -- (0.997,4.081) -- cycle
    (0.590,4.601) -- (0.551,4.518) -- (0.523,4.517) -- (0.541,4.585) -- (0.590,4.601) -- cycle
    (0.434,4.960) -- (0.392,4.959) -- (0.296,4.865) -- (0.256,4.874) -- (0.264,4.927) -- (0.348,4.943) -- (0.492,5.053) -- (0.434,4.960) -- cycle
    (0.638,5.198) -- (0.629,5.158) -- (0.592,5.175) -- (0.619,5.217) -- (0.638,5.198) -- cycle
    (0.560,5.305) -- (0.591,5.262) -- (0.551,5.210) -- (0.496,5.272) -- (0.560,5.305) -- cycle
    (0.669,6.359) -- (0.681,6.251) -- (0.606,6.174) -- (0.684,6.195) -- (0.677,6.156) -- (0.593,6.127) -- (0.591,6.024) -- (0.542,5.958) -- (0.464,5.936) -- (0.470,5.901) -- (0.299,5.793) -- (0.192,5.459) -- (0.196,5.350) -- (0.000,5.139) -- (0.066,5.309) -- (0.133,5.354) -- (0.107,5.474) -- (0.157,5.673) -- (0.075,5.745) -- (0.206,5.838) -- (0.228,5.815) -- (0.262,5.832) -- (0.260,5.789) -- (0.293,5.810) -- (0.290,5.873) -- (0.255,5.899) -- (0.316,5.904) -- (0.292,5.944) -- (0.333,5.977) -- (0.267,6.045) -- (0.297,6.156) -- (0.380,6.231) -- (0.528,6.282) -- (0.639,6.382) -- (0.669,6.359) -- cycle
    (1.880,6.436) -- (1.906,6.425) -- (1.863,6.332) -- (1.895,6.303) -- (1.871,6.224) -- (1.521,5.933) -- (1.508,5.853) -- (1.461,5.844) -- (1.600,5.847) -- (1.412,5.592) -- (1.498,5.668) -- (1.562,5.659) -- (1.767,5.752) -- (1.890,5.705) -- (1.959,5.735) -- (2.085,5.707) -- (2.328,5.712) -- (2.372,5.661) -- (2.395,5.554) -- (2.289,5.398) -- (2.214,5.118) -- (2.131,5.027) -- (2.081,4.885) -- (2.001,4.810) -- (1.869,4.805) -- (1.775,4.736) -- (1.767,4.686) -- (1.730,4.671) -- (1.779,4.617) -- (1.733,4.580) -- (1.623,4.573) -- (1.649,4.618) -- (1.565,4.637) -- (1.429,4.553) -- (1.418,4.488) -- (1.354,4.505) -- (1.304,4.480) -- (1.191,4.506) -- (1.134,4.593) -- (1.069,4.366) -- (1.099,4.326) -- (1.075,4.273) -- (0.979,4.331) -- (0.992,4.382) -- (0.922,4.382) -- (0.917,4.327) -- (0.846,4.239) -- (0.806,3.973) -- (0.707,3.981) -- (0.693,4.041) -- (0.731,4.091) -- (0.763,4.263) -- (0.727,4.229) -- (0.718,4.271) -- (0.743,4.319) -- (0.760,4.277) -- (0.792,4.312) -- (0.740,4.469) -- (0.808,4.568) -- (0.790,4.595) -- (0.649,4.360) -- (0.615,4.349) -- (0.608,4.256) -- (0.492,4.210) -- (0.480,4.239) -- (0.514,4.275) -- (0.487,4.329) -- (0.425,4.281) -- (0.403,4.304) -- (0.444,4.419) -- (0.571,4.480) -- (0.602,4.449) -- (0.634,4.481) -- (0.659,4.541) -- (0.746,4.607) -- (0.759,4.674) -- (0.724,4.675) -- (0.791,4.705) -- (0.787,4.743) -- (0.833,4.762) -- (0.810,4.781) -- (0.884,4.840) -- (0.810,4.832) -- (0.722,4.736) -- (0.699,4.766) -- (0.509,4.705) -- (0.460,4.738) -- (0.462,4.786) -- (0.547,4.772) -- (0.591,4.823) -- (0.531,4.868) -- (0.532,4.912) -- (0.465,4.871) -- (0.541,4.942) -- (0.533,4.974) -- (0.616,5.006) -- (0.661,4.991) -- (0.670,5.011) -- (0.593,5.024) -- (0.577,5.059) -- (0.688,5.097) -- (0.714,5.075) -- (0.758,5.144) -- (0.717,5.148) -- (0.703,5.182) -- (0.755,5.261) -- (0.820,5.234) -- (0.766,5.295) -- (0.823,5.349) -- (0.790,5.373) -- (0.675,5.265) -- (0.685,5.342) -- (0.720,5.377) -- (0.686,5.402) -- (0.651,5.344) -- (0.630,5.396) -- (0.588,5.352) -- (0.486,5.526) -- (0.428,5.516) -- (0.389,5.613) -- (0.414,5.653) -- (0.445,5.645) -- (0.458,5.723) -- (0.508,5.707) -- (0.486,5.678) -- (0.526,5.644) -- (0.559,5.692) -- (0.540,5.773) -- (0.563,5.756) -- (0.599,5.803) -- (0.658,5.692) -- (0.635,5.561) -- (0.666,5.557) -- (0.725,5.687) -- (0.704,5.500) -- (0.770,5.444) -- (0.797,5.463) -- (0.835,5.517) -- (0.753,5.512) -- (0.769,5.678) -- (0.824,5.671) -- (0.780,5.722) -- (0.819,5.783) -- (0.783,5.790) -- (0.791,5.884) -- (0.874,5.932) -- (0.914,5.885) -- (0.958,5.914) -- (0.978,5.971) -- (0.909,5.955) -- (0.950,6.060) -- (1.002,6.044) -- (0.968,6.159) -- (1.062,6.172) -- (1.049,6.260) -- (1.098,6.309) -- (1.090,6.380) -- (1.117,6.378) -- (1.144,6.447) -- (1.243,6.393) -- (1.397,6.382) -- (1.427,6.355) -- (1.517,6.408) -- (1.579,6.378) -- (1.659,6.417) -- (1.769,6.397) -- (1.766,6.451) -- (1.824,6.436) -- (1.869,6.469) -- (1.880,6.436) -- cycle
    (1.896,6.839) -- (1.953,6.823) -- (1.962,6.739) -- (1.997,6.754) -- (1.952,6.666) -- (2.030,6.672) -- (2.047,6.616) -- (1.997,6.619) -- (1.931,6.487) -- (1.904,6.551) -- (1.802,6.524) -- (1.746,6.602) -- (1.786,6.658) -- (1.785,6.796) -- (1.864,6.803) -- (1.896,6.839) -- cycle
    (1.975,6.960) -- (1.961,6.913) -- (1.993,6.874) -- (2.043,6.861) -- (2.064,6.919) -- (2.101,6.915) -- (2.093,6.890) -- (2.138,6.923) -- (2.160,6.898) -- (2.047,6.840) -- (2.096,6.742) -- (2.056,6.765) -- (2.035,6.743) -- (2.034,6.800) -- (1.981,6.776) -- (1.963,6.875) -- (1.908,6.895) -- (1.892,6.940) -- (1.964,6.978) -- (1.975,6.960) -- cycle
    (2.698,8.061) -- (2.750,8.040) -- (2.711,7.913) -- (2.749,7.870) -- (2.703,7.844) -- (2.672,7.873) -- (2.657,7.790) -- (2.617,7.771) -- (2.652,7.768) -- (2.627,7.724) -- (2.700,7.723) -- (2.721,7.684) -- (2.640,7.667) -- (2.601,7.577) -- (2.631,7.585) -- (2.668,7.529) -- (2.592,7.502) -- (2.614,7.426) -- (2.574,7.418) -- (2.553,7.315) -- (2.522,7.364) -- (2.527,7.537) -- (2.411,7.600) -- (2.386,7.678) -- (2.477,7.670) -- (2.499,7.732) -- (2.430,7.775) -- (2.441,7.835) -- (2.547,7.922) -- (2.582,7.867) -- (2.616,7.977) -- (2.669,7.980) -- (2.698,8.061) -- cycle;
  \filldraw[region, fill=r1]
    (1.949,4.798) -- (1.973,4.796) -- (1.972,4.728) -- (2.060,4.671) -- (1.966,4.603) -- (1.901,4.619) -- (1.815,4.516) -- (1.771,4.511) -- (1.771,4.468) -- (1.877,4.439) -- (1.961,4.522) -- (2.036,4.530) -- (2.070,4.474) -- (2.244,4.408) -- (2.320,4.271) -- (2.394,4.215) -- (2.240,4.157) -- (2.217,4.176) -- (2.197,4.139) -- (2.217,4.056) -- (2.063,3.963) -- (2.032,3.894) -- (1.918,3.787) -- (1.853,3.772) -- (1.841,3.712) -- (1.632,3.721) -- (1.608,3.624) -- (1.572,3.647) -- (1.452,3.565) -- (1.376,3.575) -- (1.274,3.662) -- (1.302,3.595) -- (1.273,3.503) -- (1.086,3.648) -- (1.038,3.612) -- (1.079,3.483) -- (1.043,3.487) -- (0.939,3.722) -- (0.983,3.773) -- (1.020,3.693) -- (0.995,3.776) -- (1.026,3.859) -- (1.196,4.116) -- (1.173,4.189) -- (1.075,4.273) -- (1.099,4.326) -- (1.069,4.366) -- (1.133,4.593) -- (1.191,4.506) -- (1.414,4.486) -- (1.429,4.553) -- (1.565,4.637) -- (1.649,4.618) -- (1.623,4.573) -- (1.733,4.580) -- (1.779,4.617) -- (1.730,4.671) -- (1.767,4.686) -- (1.775,4.736) -- (1.949,4.798) -- cycle;
  \filldraw[region, fill=r2]
    (2.400,4.192) -- (2.467,4.205) -- (2.442,4.167) -- (2.504,3.845) -- (2.661,3.463) -- (2.917,3.352) -- (2.999,3.196) -- (3.054,3.159) -- (2.925,3.138) -- (2.883,3.020) -- (2.788,3.021) -- (2.769,2.904) -- (2.719,2.871) -- (2.684,2.881) -- (2.625,2.985) -- (2.496,2.954) -- (2.434,2.992) -- (2.377,2.985) -- (2.357,3.059) -- (2.256,3.159) -- (2.163,3.176) -- (2.147,3.251) -- (2.231,3.366) -- (2.148,3.496) -- (2.168,3.578) -- (2.128,3.621) -- (2.071,3.585) -- (2.041,3.633) -- (2.066,3.675) -- (2.045,3.714) -- (2.095,3.759) -- (2.091,3.792) -- (2.012,3.844) -- (2.028,3.921) -- (2.217,4.056) -- (2.197,4.139) -- (2.217,4.177) -- (2.240,4.157) -- (2.384,4.218) -- (2.400,4.192) -- cycle;
  \filldraw[region, fill=r2]
    (2.203,3.173) -- (2.340,3.078) -- (2.376,2.986) -- (2.434,2.992) -- (2.496,2.954) -- (2.625,2.985) -- (2.718,2.871) -- (2.769,2.904) -- (2.788,3.021) -- (2.883,3.020) -- (2.904,3.048) -- (2.925,3.138) -- (3.028,3.156) -- (3.129,3.097) -- (3.082,3.045) -- (3.085,2.990) -- (3.245,2.702) -- (3.226,2.683) -- (3.239,2.712) -- (3.195,2.737) -- (3.133,2.726) -- (3.052,2.802) -- (2.951,2.768) -- (3.051,2.788) -- (3.192,2.649) -- (3.252,2.627) -- (3.295,2.540) -- (3.242,2.448) -- (3.169,2.458) -- (3.146,2.492) -- (2.976,2.512) -- (2.905,2.499) -- (2.884,2.448) -- (2.842,2.435) -- (2.827,2.590) -- (2.777,2.592) -- (2.772,2.619) -- (2.732,2.609) -- (2.705,2.492) -- (2.629,2.473) -- (2.560,2.502) -- (2.519,2.460) -- (2.464,2.482) -- (2.480,2.504) -- (2.448,2.532) -- (2.438,2.496) -- (2.387,2.484) -- (2.340,2.552) -- (2.397,2.616) -- (2.325,2.663) -- (2.277,2.754) -- (2.224,2.766) -- (2.280,2.876) -- (2.257,2.907) -- (2.197,2.900) -- (2.254,3.049) -- (2.174,3.139) -- (2.203,3.173) -- cycle;
  \filldraw[region, fill=r2]
    (2.019,3.837) -- (2.093,3.789) -- (2.045,3.714) -- (2.056,3.598) -- (2.128,3.621) -- (2.168,3.578) -- (2.148,3.496) -- (2.231,3.366) -- (2.147,3.251) -- (2.189,3.167) -- (2.174,3.139) -- (2.254,3.051) -- (2.193,2.936) -- (2.199,2.896) -- (2.257,2.907) -- (2.280,2.876) -- (2.224,2.766) -- (2.277,2.754) -- (2.325,2.663) -- (2.397,2.616) -- (2.340,2.552) -- (2.401,2.479) -- (2.377,2.469) -- (2.392,2.438) -- (2.309,2.420) -- (2.264,2.374) -- (2.164,2.418) -- (2.170,2.533) -- (2.083,2.539) -- (2.061,2.583) -- (2.022,2.571) -- (1.981,2.627) -- (1.954,2.643) -- (1.937,2.615) -- (1.904,2.676) -- (1.845,2.684) -- (1.890,2.790) -- (1.834,2.815) -- (1.837,2.933) -- (1.911,2.952) -- (1.894,3.015) -- (1.945,3.078) -- (1.921,3.134) -- (1.878,3.095) -- (1.847,3.135) -- (1.801,3.037) -- (1.764,3.059) -- (1.749,3.117) -- (1.775,3.168) -- (1.757,3.131) -- (1.722,3.133) -- (1.590,3.375) -- (1.711,3.652) -- (1.758,3.656) -- (1.737,3.675) -- (1.780,3.701) -- (1.840,3.689) -- (1.856,3.775) -- (1.900,3.774) -- (2.007,3.868) -- (2.019,3.837) -- cycle;
  \filldraw[region, fill=r2]
    (1.875,2.770) -- (1.886,2.737) -- (1.845,2.684) -- (1.904,2.676) -- (1.937,2.615) -- (1.954,2.643) -- (1.981,2.627) -- (2.022,2.571) -- (2.061,2.583) -- (2.083,2.539) -- (2.170,2.533) -- (2.164,2.418) -- (2.216,2.349) -- (2.122,2.283) -- (2.105,2.226) -- (2.043,2.221) -- (2.045,2.163) -- (2.013,2.156) -- (2.053,2.107) -- (2.033,2.084) -- (1.848,2.077) -- (1.867,2.041) -- (1.841,2.048) -- (1.786,1.946) -- (1.782,1.874) -- (1.614,1.820) -- (1.581,1.781) -- (1.518,1.789) -- (1.417,1.741) -- (1.384,1.758) -- (1.357,1.716) -- (1.313,1.745) -- (1.369,1.882) -- (1.337,1.974) -- (1.372,2.049) -- (1.332,2.120) -- (1.337,2.190) -- (1.231,2.185) -- (1.183,2.144) -- (1.192,2.113) -- (1.162,2.101) -- (1.131,2.136) -- (1.037,2.084) -- (1.076,2.166) -- (1.253,2.306) -- (1.242,2.365) -- (1.100,2.500) -- (1.157,2.521) -- (1.146,2.587) -- (1.224,2.591) -- (1.290,2.575) -- (1.323,2.487) -- (1.406,2.502) -- (1.376,2.436) -- (1.479,2.482) -- (1.470,2.516) -- (1.584,2.476) -- (1.715,2.522) -- (1.809,2.437) -- (1.753,2.538) -- (1.844,2.586) -- (1.809,2.666) -- (1.875,2.770) -- cycle;
  \filldraw[region, fill=r3]
    (1.683,1.843) -- (1.799,1.768) -- (1.833,1.705) -- (1.782,1.614) -- (1.811,1.500) -- (1.855,1.441) -- (1.898,1.462) -- (1.997,1.386) -- (1.983,1.310) -- (2.029,1.258) -- (1.973,1.199) -- (1.786,1.133) -- (1.749,1.064) -- (1.577,1.070) -- (1.442,1.234) -- (1.308,1.181) -- (1.222,1.201) -- (1.265,1.267) -- (1.325,1.267) -- (1.230,1.283) -- (1.207,1.352) -- (1.071,1.326) -- (1.067,1.256) -- (0.987,1.269) -- (0.952,1.234) -- (0.888,1.254) -- (0.829,1.323) -- (0.784,1.314) -- (0.809,1.335) -- (0.778,1.354) -- (0.867,1.372) -- (0.868,1.419) -- (0.735,1.433) -- (0.742,1.470) -- (0.767,1.455) -- (0.886,1.519) -- (0.894,1.563) -- (0.999,1.554) -- (1.055,1.622) -- (1.153,1.632) -- (1.313,1.745) -- (1.357,1.716) -- (1.384,1.758) -- (1.417,1.741) -- (1.518,1.789) -- (1.581,1.781) -- (1.683,1.843) -- cycle;
  \filldraw[region, fill=r3]
    (2.317,2.408) -- (2.379,2.372) -- (2.406,2.275) -- (2.320,2.164) -- (2.438,2.029) -- (2.395,1.940) -- (2.478,1.895) -- (2.446,1.815) -- (2.466,1.779) -- (2.369,1.744) -- (2.375,1.710) -- (2.529,1.605) -- (2.603,1.634) -- (2.640,1.586) -- (2.608,1.574) -- (2.629,1.514) -- (2.568,1.489) -- (2.517,1.507) -- (2.367,1.408) -- (2.302,1.446) -- (2.300,1.335) -- (2.273,1.312) -- (2.304,1.264) -- (2.183,1.239) -- (2.163,1.086) -- (2.072,1.078) -- (2.056,1.098) -- (2.107,1.126) -- (2.069,1.140) -- (2.041,1.208) -- (1.996,1.211) -- (2.029,1.258) -- (1.983,1.310) -- (2.001,1.381) -- (1.898,1.462) -- (1.838,1.458) -- (1.782,1.614) -- (1.833,1.705) -- (1.799,1.768) -- (1.696,1.840) -- (1.782,1.874) -- (1.786,1.946) -- (1.841,2.048) -- (1.867,2.041) -- (1.848,2.077) -- (2.033,2.084) -- (2.053,2.107) -- (2.013,2.156) -- (2.045,2.163) -- (2.043,2.221) -- (2.129,2.247) -- (2.122,2.283) -- (2.216,2.349) -- (2.190,2.376) -- (2.264,2.374) -- (2.317,2.408) -- cycle;
  \filldraw[region, fill=r3]
    (3.315,2.125) -- (3.299,2.081) -- (3.219,2.053) -- (3.195,1.989) -- (3.169,2.019) -- (3.114,1.974) -- (3.037,1.996) -- (3.011,1.929) -- (3.048,1.862) -- (2.978,1.787) -- (2.986,1.668) -- (2.922,1.646) -- (2.944,1.592) -- (2.913,1.561) -- (2.913,1.485) -- (2.871,1.490) -- (2.847,1.432) -- (2.731,1.419) -- (2.728,1.517) -- (2.626,1.495) -- (2.608,1.574) -- (2.640,1.585) -- (2.603,1.634) -- (2.529,1.605) -- (2.375,1.710) -- (2.369,1.744) -- (2.466,1.779) -- (2.446,1.815) -- (2.478,1.895) -- (2.390,1.950) -- (2.438,2.029) -- (2.320,2.164) -- (2.406,2.275) -- (2.379,2.372) -- (2.303,2.413) -- (2.392,2.438) -- (2.377,2.469) -- (2.438,2.496) -- (2.448,2.532) -- (2.480,2.504) -- (2.464,2.482) -- (2.519,2.460) -- (2.560,2.502) -- (2.629,2.473) -- (2.705,2.492) -- (2.750,2.619) -- (2.827,2.590) -- (2.844,2.435) -- (2.905,2.499) -- (2.976,2.512) -- (3.146,2.492) -- (3.169,2.458) -- (3.242,2.448) -- (3.295,2.540) -- (3.349,2.409) -- (3.344,2.336) -- (3.216,2.194) -- (3.315,2.125) -- cycle;
  \filldraw[region, fill=r3]
    (3.315,2.125) -- (3.390,2.156) -- (3.432,2.254) -- (3.630,2.265) -- (3.774,2.236) -- (3.957,2.096) -- (4.000,1.909) -- (3.934,1.614) -- (3.818,1.490) -- (3.797,1.507) -- (3.803,1.424) -- (3.734,1.369) -- (3.621,1.351) -- (3.655,1.342) -- (3.668,1.256) -- (3.587,1.177) -- (3.434,1.157) -- (3.406,1.117) -- (3.283,1.166) -- (3.250,1.283) -- (3.193,1.206) -- (3.033,1.178) -- (2.974,1.264) -- (2.933,1.240) -- (2.877,1.287) -- (2.836,1.259) -- (2.781,1.315) -- (2.746,1.305) -- (2.693,1.388) -- (2.767,1.439) -- (2.847,1.432) -- (2.871,1.490) -- (2.913,1.485) -- (2.913,1.561) -- (2.944,1.592) -- (2.922,1.646) -- (2.986,1.668) -- (2.978,1.787) -- (3.048,1.862) -- (3.011,1.929) -- (3.037,1.996) -- (3.114,1.974) -- (3.169,2.019) -- (3.195,1.989) -- (3.219,2.053) -- (3.299,2.081) -- (3.315,2.125) -- cycle;
  \filldraw[region, fill=r4]
    (3.109,1.140) -- (3.101,1.198) -- (3.193,1.206) -- (3.250,1.283) -- (3.283,1.166) -- (3.341,1.109) -- (3.280,1.103) -- (3.247,1.054) -- (3.180,1.093) -- (3.105,1.059) -- (3.109,1.140) -- cycle;
  \filldraw[region, fill=r5]
    (3.483,1.151) -- (3.641,1.107) -- (3.675,1.056) -- (3.891,1.096) -- (3.887,1.043) -- (3.860,1.040) -- (3.884,0.935) -- (3.856,0.888) -- (3.717,0.833) -- (3.692,0.730) -- (3.598,0.741) -- (3.545,0.693) -- (3.408,0.648) -- (3.357,0.588) -- (3.155,0.646) -- (2.968,0.640) -- (3.012,0.695) -- (2.974,0.689) -- (2.970,0.738) -- (3.061,0.816) -- (3.047,0.857) -- (2.954,0.844) -- (2.961,0.911) -- (2.901,0.991) -- (2.971,1.033) -- (2.951,1.055) -- (2.997,1.092) -- (3.030,1.066) -- (3.088,1.138) -- (3.104,1.059) -- (3.180,1.093) -- (3.242,1.054) -- (3.280,1.103) -- (3.340,1.108) -- (3.334,1.133) -- (3.406,1.117) -- (3.483,1.151) -- cycle;
  \filldraw[region, fill=r5]
    (2.704,1.523) -- (2.731,1.513) -- (2.730,1.395) -- (2.693,1.388) -- (2.746,1.305) -- (2.781,1.315) -- (2.836,1.259) -- (2.877,1.287) -- (2.933,1.240) -- (2.962,1.269) -- (3.020,1.182) -- (3.109,1.168) -- (3.047,1.079) -- (2.997,1.092) -- (2.950,1.054) -- (2.971,1.033) -- (2.902,0.994) -- (2.961,0.911) -- (2.954,0.844) -- (3.060,0.841) -- (3.042,0.778) -- (2.996,0.774) -- (2.968,0.733) -- (2.975,0.689) -- (3.012,0.698) -- (3.000,0.654) -- (2.868,0.565) -- (2.629,0.645) -- (2.622,0.604) -- (2.740,0.541) -- (2.654,0.458) -- (2.495,0.516) -- (2.503,0.550) -- (2.439,0.571) -- (2.342,0.547) -- (2.324,0.467) -- (2.276,0.453) -- (2.100,0.496) -- (2.083,0.407) -- (2.074,0.456) -- (1.974,0.534) -- (1.999,0.608) -- (1.955,0.656) -- (2.019,0.685) -- (1.971,0.776) -- (2.075,0.852) -- (2.131,0.942) -- (2.112,0.979) -- (2.168,1.043) -- (2.183,1.239) -- (2.304,1.264) -- (2.273,1.312) -- (2.300,1.335) -- (2.302,1.446) -- (2.367,1.408) -- (2.517,1.507) -- (2.665,1.494) -- (2.704,1.523) -- cycle;
  \filldraw[region, fill=r5]
    (0.231,0.000) -- (0.172,0.000) -- (0.208,0.018) -- (0.208,0.062) -- (0.255,0.077) -- (0.278,0.046) -- (0.231,0.000) -- cycle
    (2.007,1.206) -- (2.041,1.208) -- (2.069,1.140) -- (2.105,1.135) -- (2.064,1.084) -- (2.173,1.081) -- (2.075,0.852) -- (1.971,0.776) -- (2.019,0.685) -- (1.955,0.656) -- (2.003,0.583) -- (1.974,0.534) -- (1.873,0.571) -- (1.633,0.486) -- (1.591,0.414) -- (1.604,0.377) -- (1.569,0.362) -- (1.602,0.330) -- (1.528,0.256) -- (1.524,0.200) -- (1.438,0.196) -- (1.374,0.264) -- (1.247,0.273) -- (1.210,0.309) -- (1.035,0.281) -- (1.024,0.303) -- (0.972,0.214) -- (0.901,0.205) -- (0.870,0.159) -- (0.854,0.205) -- (0.842,0.085) -- (0.775,0.030) -- (0.723,0.120) -- (0.632,0.159) -- (0.612,0.113) -- (0.550,0.086) -- (0.516,0.121) -- (0.555,0.198) -- (0.615,0.233) -- (0.669,0.215) -- (0.759,0.276) -- (0.810,0.362) -- (0.857,0.383) -- (0.885,0.483) -- (0.984,0.502) -- (0.998,0.551) -- (1.097,0.633) -- (1.118,0.808) -- (1.223,0.786) -- (1.267,0.836) -- (1.261,0.926) -- (1.471,0.965) -- (1.650,0.913) -- (1.804,0.925) -- (1.844,1.061) -- (2.007,1.206) -- cycle;

  \node[letterlbl] at (1.368,5.474) {P};
  \node[letterlbl] at (1.604,4.124) {N};
  \node[letterlbl] at (2.454,3.490) {F};
  \node[letterlbl] at (2.742,2.770) {M};
  \node[letterlbl] at (1.899,2.781) {G};
  \node[letterlbl] at (1.649,2.242) {D};
  \node[letterlbl] at (1.458,1.438) {K};
  \node[letterlbl] at (2.152,1.753) {E};
  \node[letterlbl] at (2.795,2.080) {B};
  \node[letterlbl] at (3.431,1.686) {A};
  \node[letterlbl] at (3.206,1.149) {C};
  \node[letterlbl] at (3.377,0.893) {J};
  \node[letterlbl] at (2.520,0.938) {H};
  \node[letterlbl] at (1.443,0.613) {L};

  \node[macrolbl, text=r1d] at (1.634,4.743) {1};
  \node[macrolbl, text=r2d] at (2.165,3.014) {2};
  \node[macrolbl, text=r3d] at (3.030,1.751) {3};
  \node[macrolbl, text=r4d] at (3.209,1.232) {4};
  \node[macrolbl, text=r5d] at (2.165,0.621) {5};
\end{tikzpicture}

%% file: figures/timeline_modes.tex
\begin{tikzpicture}[x=1.32cm, y=1.2cm, font=\small]

  \definecolor{cbatch}{HTML}{E89A28}   
  \definecolor{cupdate}{HTML}{2898E8}  

  \fill[black!8] (0,   1.35) rectangle (1.5, -3.28);   
  \fill[black!3] (1.5, 1.35) rectangle (7.5, -3.28);   

  \foreach \x in {1.5, 3.5, 5.5} {
    \draw[black!20, densely dashed, line width=0.4pt] (\x, 1.30) -- (\x, -3.23);
  }
  \draw[black!20, line width=0.4pt] (7.5, 1.30) -- (7.5, -3.23);

  \node[font=\scriptsize, black!60, anchor=south] at (0.75, 1.30) {Training Set};
  \node[font=\scriptsize, black!60, anchor=south] at (4.50, 1.30) {Test Set};

  \draw[black!18, line width=0.35pt] (1.5, 0.92) -- (7.5, 0.92);

  \node[font=\scriptsize, black!45, anchor=south] at (2.5, 0.65) {Month $1$};
  \node[font=\scriptsize, black!45, anchor=south] at (4.5, 0.65) {Month $2$};
  \node[font=\scriptsize, black!45, anchor=south] at (6.5, 0.65) {Month $3$};

  \node[anchor=east, font=\small] at (-0.12,  0.00) {Repeated Batch};
  \node[anchor=east, font=\small] at (-0.12, -1.40) {Repeated Online};
  \node[anchor=east, font=\small] at (-0.12, -2.80) {Online};

  \draw[->, black!35, line width=0.5pt] (0, 0) -- (7.80, 0);

  \fill[black!12] (0, -0.18) rectangle (1.5, 0.18);
  \draw[black!25, line width=0.3pt] (0, -0.18) rectangle (1.5, 0.18);

  \foreach \xa/\xb in {1.5/3.5, 3.5/5.5, 5.5/7.5} {
    \fill[cbatch!22] (\xa, -0.18) rectangle (\xb,  0.18);
    \draw[cbatch!55, line width=0.3pt] (\xa, -0.18) rectangle (\xb, 0.18);
  }

  \foreach \x in {1.5, 3.5, 5.5} {
    \fill[cbatch]                      (\x, 0) circle (3.5pt);
    \draw[black!60, line width=0.35pt] (\x, 0) circle (3.5pt);
  }

  \draw[->, black!35, line width=0.5pt] (0, -1.4) -- (7.80, -1.4);

  \fill[black!12] (0, -1.58) rectangle (1.5, -1.22);
  \draw[black!25, line width=0.3pt] (0, -1.58) rectangle (1.5, -1.22);

  \foreach \xstart in {1.5, 3.5, 5.5} {
    \fill[cbatch]                      (\xstart, -1.4) circle (3.5pt);
    \draw[black!60, line width=0.35pt] (\xstart, -1.4) circle (3.5pt);
    \foreach \k in {1,...,9} {
      \pgfmathsetmacro\xpos{\xstart + \k * 0.2}
      \draw[cupdate, line width=0.55pt] (\xpos, -1.4) -- (\xpos, -1.24);
    }
  }

  \draw[->, black!35, line width=0.5pt] (0, -2.8) -- (7.80, -2.8);

  \fill[black!12] (0, -2.98) rectangle (1.5, -2.62);
  \draw[black!25, line width=0.3pt] (0, -2.98) rectangle (1.5, -2.62);

  \fill[cbatch]                      (1.5, -2.8) circle (3.5pt);
  \draw[black!60, line width=0.35pt] (1.5, -2.8) circle (3.5pt);

  \foreach \k in {1,...,29} {
    \pgfmathsetmacro\xpos{1.5 + \k * 0.2}
    \draw[cupdate, line width=0.55pt] (\xpos, -2.8) -- (\xpos, -2.64);
  }

  \fill[cbatch]                      (0.40, -3.55) circle (3pt);
  \draw[black!60, line width=0.35pt] (0.40, -3.55) circle (3pt);
  \node[anchor=west, font=\scriptsize] at (0.60, -3.55)
    {batch fit};

  \fill[cbatch!22]                   (2.35, -3.64) rectangle (2.75, -3.46);
  \draw[cbatch!55, line width=0.3pt] (2.35, -3.64) rectangle (2.75, -3.46);
  \node[anchor=west, font=\scriptsize] at (2.85, -3.55) {parameters fixed};

  \draw[cupdate, line width=0.55pt] (5.60, -3.55) -- (5.60, -3.37);
  \node[anchor=west, font=\scriptsize] at (5.75, -3.55) {online update};

\end{tikzpicture}

%% file: tables/tab_forecast_eval.tex
\begingroup
\setlength{\tabcolsep}{9pt}
\begin{tabular}{ll@{\hspace{1.0em}}cccc}
\toprule
Model & Scheme & ES & VS$_{0.5}$ & DSS & LS \\
\midrule
M+GC & Online & 1.3719 & 2.3214 & -0.8911 & 4.1381 \\
& Repeated Online & \cellcolor{green!25}1.2584 & \cellcolor{green!25}2.1003 & \cellcolor{green!25}-1.4932 & \cellcolor{green!25}3.8366 \\
& Repeated Batch & 1.3593 & 2.3350 & -0.4302 & 4.3652 \\
\addlinespace[6pt]
M+VC & Online & 1.3719 & 2.3215 & -0.8905 & 4.1377 \\
& Repeated Online & \cellcolor{green!25}1.2584 & \cellcolor{green!25}2.1004 & \cellcolor{green!25}-1.4935 & \cellcolor{green!25}3.8358 \\
& Repeated Batch & 1.3593 & 2.3350 & -0.4307 & 4.3624 \\
\addlinespace[6pt]
M+CVC & Online & 1.3716 & 2.3179 & -0.9630 & 4.1010 \\
& Repeated Online & \cellcolor{green!25}\underline{1.2580} & \cellcolor{green!25}\underline{2.0983} & \cellcolor{green!25}\underline{-1.5580} & \cellcolor{green!25}\underline{3.8028} \\
& Repeated Batch & 1.3591 & 2.3344 & -0.4681 & 4.3432 \\
\bottomrule
\end{tabular}
\endgroup

%% file: tables/tab_mcs.tex
\begingroup
\setlength{\tabcolsep}{9pt}
\begin{tabular}{ll@{\hspace{1.0em}}rrrr}
\toprule
Model & Scheme & ES & VS$_{0.5}$ & DSS & LS \\
\midrule
M+GC & Online & 0.000 & 0.000 & 0.000 & 0.000 \\
& Repeated Online & 0.000 & 0.068 & 0.000 & 0.000 \\
& Repeated Batch & 0.000 & 0.000 & 0.000 & 0.000 \\
\addlinespace[6pt]
M+VC & Online & 0.000 & 0.000 & 0.000 & 0.000 \\
& Repeated Online & 0.000 & 0.058 & 0.000 & 0.000 \\
& Repeated Batch & 0.000 & 0.000 & 0.000 & 0.000 \\
\addlinespace[6pt]
M+CVC & Online & 0.000 & 0.000 & 0.000 & 0.000 \\
& Repeated Online & \textbf{1.000} & \textbf{1.000} & \textbf{1.000} & \textbf{1.000} \\
& Repeated Batch & 0.000 & 0.000 & 0.000 & 0.000 \\
\bottomrule
\end{tabular}
\endgroup

%% file: tables/table_vine_time_ols_avg.tex
\begin{tabular}{lrrrrr}
\toprule
 & $\gamma=0$ & $\gamma=1/400$ & $\gamma=1/200$ & $\gamma=1/100$ & $\gamma=1/50$ \\
\midrule
Batch & 1.669 & 1.307 & 1.048 & 0.879 & 0.800 \\
Online & 0.500 & 0.504 & 0.504 & 0.505 & 0.538 \\
\bottomrule
\end{tabular}

%% file: tables/tab_computation_times.tex
\begin{tabular}{llrrrrrrr}
\toprule
& & \multicolumn{3}{c}{Fit} & \multicolumn{3}{c}{Update} & \\
\cmidrule(lr){3-5}\cmidrule(lr){6-8}
Model & Scheme & $N$ & Mean (s) & Total (s) & $N$ & Mean (s) & Total (s) & Total (s) \\
\midrule
M+GC & Online & 1 & 12.94 & 12.94 & 15935 & 0.62 & 9900.67 & 9913.61 \\
M+GC & Repeated Online & 12 & 9.49 & 113.87 & 15924 & 0.43 & 6887.93 & 7001.80 \\
M+GC & Repeated Batch & 12 & 9.67 & 116.03 & -- & -- & -- & 116.03 \\
\addlinespace[6pt]
M+VC & Online & 1 & 31.15 & 31.15 & 15935 & 0.60 & 9554.70 & 9585.85 \\
M+VC & Repeated Online & 12 & 29.39 & 352.74 & 15924 & 0.42 & 6629.05 & 6981.78 \\
M+VC & Repeated Batch & 12 & 29.77 & 357.21 & -- & -- & -- & 357.21 \\
\addlinespace[6pt]
M+CVC & Online & 1 & 215.73 & 215.73 & 15935 & 2.69 & 42808.15 & 43023.88 \\
M+CVC & Repeated Online & 12 & 202.62 & 2431.45 & 15924 & 1.85 & 29481.10 & 31912.55 \\
M+CVC & Repeated Batch & 12 & 188.71 & 2264.54 & -- & -- & -- & 2264.54 \\
\bottomrule
\end{tabular}

%% file: tables/tab_mcs_tr.tex
\begingroup
\setlength{\tabcolsep}{9pt}
\begin{tabular}{ll@{\hspace{1.0em}}rrrr}
\toprule
Model & Scheme & ES & VS$_{0.5}$ & DSS & LS \\
\midrule
M+GC & Online & 0.000 & 0.000 & 0.000 & 0.000 \\
& Repeated Online & 0.000 & 0.068 & 0.000 & 0.000 \\
& Repeated Batch & 0.000 & 0.000 & 0.000 & 0.000 \\
\addlinespace[6pt]
M+VC & Online & 0.000 & 0.000 & 0.000 & 0.000 \\
& Repeated Online & 0.000 & 0.057 & 0.000 & 0.000 \\
& Repeated Batch & 0.000 & 0.000 & 0.000 & 0.000 \\
\addlinespace[6pt]
M+CVC & Online & 0.000 & 0.000 & 0.000 & 0.000 \\
& Repeated Online & \textbf{1.000} & \textbf{1.000} & \textbf{1.000} & \textbf{1.000} \\
& Repeated Batch & 0.000 & 0.000 & 0.000 & 0.000 \\
\bottomrule
\end{tabular}
\endgroup

%% file: tables/tab_dm_vs_best.tex
\begingroup
\setlength{\tabcolsep}{9pt}
\begin{tabular}{ll@{\hspace{1.0em}}rrrr}
\toprule
Model & Scheme & ES & VS$_{0.5}$ & DSS & LS \\
\midrule
M+GC & Online & $12.488$\rlap{$^{***}$} & $12.093$\rlap{$^{***}$} & $12.172$\rlap{$^{***}$} & $12.344$\rlap{$^{***}$} \\
& Repeated Online & $4.507$\rlap{$^{***}$} & $1.784$\rlap{$^{*}$} & $4.060$\rlap{$^{***}$} & $4.274$\rlap{$^{***}$} \\
& Repeated Batch & $9.854$\rlap{$^{***}$} & $11.687$\rlap{$^{***}$} & $15.381$\rlap{$^{***}$} & $15.192$\rlap{$^{***}$} \\
\addlinespace[6pt]
M+VC & Online & $12.487$\rlap{$^{***}$} & $12.096$\rlap{$^{***}$} & $12.187$\rlap{$^{***}$} & $12.355$\rlap{$^{***}$} \\
& Repeated Online & $4.397$\rlap{$^{***}$} & $1.814$\rlap{$^{*}$} & $4.040$\rlap{$^{***}$} & $4.127$\rlap{$^{***}$} \\
& Repeated Batch & $9.853$\rlap{$^{***}$} & $11.688$\rlap{$^{***}$} & $15.382$\rlap{$^{***}$} & $15.537$\rlap{$^{***}$} \\
\addlinespace[6pt]
M+CVC & Online & $12.442$\rlap{$^{***}$} & $12.059$\rlap{$^{***}$} & $11.909$\rlap{$^{***}$} & $12.000$\rlap{$^{***}$} \\
& Repeated Batch & $9.827$\rlap{$^{***}$} & $11.646$\rlap{$^{***}$} & $14.839$\rlap{$^{***}$} & $14.934$\rlap{$^{***}$} \\
\bottomrule
\end{tabular}
\endgroup

%% file: tables/tab_dm_within_scheme.tex
\begingroup
\setlength{\tabcolsep}{9pt}
\begin{tabular}{ll@{\hspace{1.0em}}rrrr}
\toprule
Scheme & M+CVC vs $\cdots$ & ES & VS$_{0.5}$ & DSS & LS \\
\midrule
Online & M+GC & $4.317$\rlap{$^{***}$} & $2.903$\rlap{$^{***}$} & $4.782$\rlap{$^{***}$} & $5.031$\rlap{$^{***}$} \\
& M+VC & $4.255$\rlap{$^{***}$} & $2.941$\rlap{$^{***}$} & $4.824$\rlap{$^{***}$} & $4.941$\rlap{$^{***}$} \\
\addlinespace[6pt]
Repeated Online & M+GC & $4.507$\rlap{$^{***}$} & $1.784$\rlap{$^{*}$} & $4.060$\rlap{$^{***}$} & $4.274$\rlap{$^{***}$} \\
& M+VC & $4.397$\rlap{$^{***}$} & $1.814$\rlap{$^{*}$} & $4.040$\rlap{$^{***}$} & $4.127$\rlap{$^{***}$} \\
\addlinespace[6pt]
Repeated Batch & M+GC & $2.347$\rlap{$^{**}$} & $0.535$ & $2.310$\rlap{$^{**}$} & $2.522$\rlap{$^{**}$} \\
& M+VC & $2.255$\rlap{$^{**}$} & $0.553$ & $2.278$\rlap{$^{**}$} & $2.311$\rlap{$^{**}$} \\
\bottomrule
\end{tabular}
\endgroup

%% file: tables/tab_dm_within_copula.tex
\begingroup
\setlength{\tabcolsep}{9pt}
\begin{tabular}{ll@{\hspace{1.0em}}rrrr}
\toprule
Model & Rep.\ Online vs $\cdots$ & ES & VS$_{0.5}$ & DSS & LS \\
\midrule
M+GC & Online & $12.448$\rlap{$^{***}$} & $12.133$\rlap{$^{***}$} & $12.155$\rlap{$^{***}$} & $12.350$\rlap{$^{***}$} \\
& Repeated Batch & $9.819$\rlap{$^{***}$} & $11.613$\rlap{$^{***}$} & $14.900$\rlap{$^{***}$} & $14.878$\rlap{$^{***}$} \\
\addlinespace[6pt]
M+VC & Online & $12.449$\rlap{$^{***}$} & $12.134$\rlap{$^{***}$} & $12.176$\rlap{$^{***}$} & $12.316$\rlap{$^{***}$} \\
& Repeated Batch & $9.819$\rlap{$^{***}$} & $11.612$\rlap{$^{***}$} & $14.903$\rlap{$^{***}$} & $15.040$\rlap{$^{***}$} \\
\addlinespace[6pt]
M+CVC & Online & $12.442$\rlap{$^{***}$} & $12.059$\rlap{$^{***}$} & $11.909$\rlap{$^{***}$} & $12.000$\rlap{$^{***}$} \\
& Repeated Batch & $9.827$\rlap{$^{***}$} & $11.646$\rlap{$^{***}$} & $14.839$\rlap{$^{***}$} & $14.934$\rlap{$^{***}$} \\
\bottomrule
\end{tabular}
\endgroup